\documentclass[camera,letterpaper,nomarginnotes,nonarrowgutter]{jpaper}
\usepackage[sort,compress]{cite}
\usepackage{amsmath,amssymb,amsfonts}
\usepackage{algorithmic}
\usepackage{graphicx}
\usepackage{textcomp}
\usepackage{xcolor}
\usepackage{fancyhdr}
\usepackage[keeplastbox]{flushend}
\usepackage{adjustbox}
\usepackage{siunitx}
\usepackage{tcolorbox}
\usepackage{pifont}
\usepackage{hyperref}
\usepackage[us,12hr]{datetime}
\usepackage[en-GB, useregional=numeric]{datetime2}
\usepackage{mathptmx} 
\usepackage{setspace}

\def\BibTeX{{\rm B\kern-.05em{\sc i\kern-.025em b}\kern-.08em
    T\kern-.1667em\lower.7ex\hbox{E}\kern-.125emX}}

\newif\ifdraft
\newif\ifrevision
\draftfalse
\revisionfalse

\newif\ifdsnadds
\dsnaddsfalse

\newif\ifiterations
\iterationsfalse

\newif\ifshepherd
\shepherdtrue

\newcommand{\dingOne}{\ding{182}}
\newcommand{\dingTwo}{\ding{183}}
\newcommand{\dingThree}{\ding{184}}
\newcommand{\dingFour}{\ding{185}}
\newcommand{\dingFive}{\ding{186}}

\newcommand{\ignore}[1]{}

\ifdraft
    \definecolor{gfored}{rgb}{0.580, 0.050, 0.211}
    \definecolor{ao}{rgb}{0.007, 0.520, 0.867}
    \definecolor{moegi}{rgb}{0.357, 0.537, 0.188}
    \definecolor{jl}{rgb}{1.0, 0.2, 0.8}
    \definecolor{brown(web)}{rgb}{0.65, 0.16, 0.16}
    \definecolor{bisque}{rgb}{1.0, 0.89, 0.77}
    \definecolor{nbs}{rgb}{0.88, 0.07, 0.37}
    \definecolor{yt}{rgb}{0.58, 0.44, 0.86}
    \definecolor{iy}{rgb}{0.0, 0.26, 0.15}
    \definecolor{st}{rgb}{0,0.68,0.67}
    
    \newcommand{\param}[1]{\textcolor{red}{\textbf{#1}}}
    \newcommand{\nrh}[0]{$HC_{first}$}
    
    \newcommand{\atatodo}[1]{\textcolor{red}{\textbf{TODO:}#1}}
    \newcommand{\atbnote}[1]{\textcolor{blue}{\textbf{NOTE:}#1}}
    \newcommand{\agycomment}[1]{\textcolor{gfored}{\textbf{[@gy:} #1\textbf{]}}}
    \definecolor{ao}{rgb}{0.007, 0.520, 0.867}
    \newcommand{\agy}[1]{\textcolor{gfored}{#1}}
    \newcommand{\atbcomment}[1]{\textcolor{ao}{\textbf{[@atb:} #1\textbf{]}}}
    \newcommand{\atbcommentside}[1]{\todo[size=\scriptsize, linecolor=orange, bordercolor=orange, backgroundcolor=white]{\textcolor{ao}{\textbf{@atb:} #1}}}
    \newcommand{\atb}[1]{\textcolor{ao}{#1}}
    \newcommand{\yctcomment}[1]{\textcolor{yt}{\textbf{[@yct:} #1\textbf{]}}}
    
    \newcommand{\majdcomment}[1]{\textcolor{mo}{\textbf{[@majd:} #1\textbf{]}}}
    
    \definecolor{mo}{rgb}{0, 0.5, 0}
    \newcommand{\majdtodo}[1]{\textcolor{red}{\textbf{TODO:}#1}}

    \newcommand{\stecomment}[1]{\textcolor{st}{\textbf{[@ste:} #1\textbf{]}}}

    \newcommand{\omcomment}[1]{\todo[size=\scriptsize, linecolor=orange, bordercolor=orange, backgroundcolor=white]{\textcolor{blue}{\textbf{@onur:} #1}}}
    \newcommand{\omcommentinline}[1]{\textcolor{blue}{\textbf{[@Onur:} #1\textbf{]}}}

    \newcommand{\bscomment}[1]{\todo[size=\scriptsize, linecolor=brown, bordercolor=brown, backgroundcolor=white]{\textcolor{blue}{\textbf{@Behzad:} #1}}}
    \newcommand{\bscommentinline}[1]{\textcolor{brown}{\textbf{[@Behzad:} #1\textbf{]}}}

    \newcommand{\dsnadd}[1]{#1}

        \newcommand{\atbcr}[2]{#2}
    \newcommand{\omcr}[2]{#2}
    \newcommand{\atbcrcomment}[2]{}
    \newcommand{\omcrcomment}[2]{}
\else
\ifrevision

    \newcommand{\omcomment}[1]{}
    \newcommand{\omcommentinline}[1]{}

    \newcommand{\nrh}[0]{$HC_{first}$}

    \newcommand{\bscomment}[1]{}
    \newcommand{\bscommentinline}[1]{}

    \newcommand{\agy}[1]{{#1}}
    \newcommand{\agycomment}[1]{}
    
    \newcommand{\atb}[1]{{#1}}
    \newcommand{\atbcomment}[1]{}
    \newcommand{\atbcommentside}[1]{}
    \newcommand{\atatodo}[1]{}
    \newcommand{\atbnote}[1]{}

    \newcommand{\majdcomment}[1]{}
    \newcommand{\majdtodo}[1]{}

    \newcommand{\yctcomment}[1]{}

    \newcommand{\stecomment}[1]{}

    \newcommand{\param}[1]{{#1}} 

    \definecolor{goodgreen}{rgb}{0.0, 0.5, 0.0}

    \newcommand{\revcommon}[1]{\textcolor{blue}{#1}}
    \newcommand{\reva}[1]{\textcolor{red}{#1}}
    \newcommand{\revb}[1]{\textcolor{goodgreen}{#1}}
    \newcommand{\revc}[1]{\textcolor{orange}{#1}}
    \newcommand{\reve}[1]{\textcolor{brown}{#1}}

    \paperwidth=\dimexpr \paperwidth + 2cm\relax
    \oddsidemargin=\dimexpr\oddsidemargin + 1cm\relax
    \evensidemargin=\dimexpr\evensidemargin + 1cm\relax
    \marginparwidth=\dimexpr \marginparwidth + 1cm\relax
    
    \usepackage[colorinlistoftodos,prependcaption,textsize=small]{todonotes}
    \usepackage{xargs}
    \newcommandx{\revlabel}[2][1=]{\todo[linecolor=blue,backgroundcolor=blue!25,bordercolor=blue,#1,size=\scriptsize]{#2}}

    \newcommand{\dsnadd}[1]{#1}
    \newcommand{\atbcr}[2]{#2}
    \newcommand{\omcr}[2]{#2}
    \newcommand{\atbcrcomment}[2]{}
    \newcommand{\omcrcomment}[2]{}
\else
\ifdsnadds

    \newcommand{\omcomment}[1]{}
    \newcommand{\omcommentinline}[1]{}

    \newcommand{\nrh}[0]{$HC_{first}$}

    \newcommand{\bscomment}[1]{}
    \newcommand{\bscommentinline}[1]{}

    \newcommand{\agy}[1]{{#1}}
    \newcommand{\agycomment}[1]{}
    
    \newcommand{\atb}[1]{{#1}}
    \newcommand{\atbcomment}[1]{}
    \newcommand{\atbcommentside}[1]{}
    \newcommand{\atatodo}[1]{}
    \newcommand{\atbnote}[1]{}

    \newcommand{\majdcomment}[1]{}
    \newcommand{\majdtodo}[1]{}

    \newcommand{\yctcomment}[1]{}

    \newcommand{\stecomment}[1]{}

    \newcommand{\param}[1]{{#1}} 

    \newcommand{\revcommon}[1]{{#1}}
    \newcommand{\reva}[1]{{#1}}
    \newcommand{\revb}[1]{{#1}}
    \newcommand{\revc}[1]{{#1}}
    
    \newcommand{\revlabel}[2][1=]{}

    \newcommand{\dsnadd}[1]{\textcolor{blue}{#1}}

    \usepackage[disable]{todonotes}
    \newcommand{\atbcr}[2]{#2}
    \newcommand{\omcr}[2]{#2}
    \newcommand{\atbcrcomment}[2]{}
    \newcommand{\omcrcomment}[2]{}

\else
\ifiterations

    \newcommand{\omcomment}[1]{}
    \newcommand{\omcommentinline}[1]{}

    \newcommand{\nrh}[0]{$HC_{first}$}

    \newcommand{\bscomment}[1]{}
    \newcommand{\bscommentinline}[1]{}

    \newcommand{\agy}[1]{{#1}}
    \newcommand{\agycomment}[1]{}
    
    \newcommand{\atb}[1]{{#1}}
    \newcommand{\atbcomment}[1]{}
    \newcommand{\atbcommentside}[1]{}
    \newcommand{\atatodo}[1]{}
    \newcommand{\atbnote}[1]{}

    \newcommand{\majdcomment}[1]{}
    \newcommand{\majdtodo}[1]{}

    \newcommand{\yctcomment}[1]{}

    \newcommand{\stecomment}[1]{}

    \newcommand{\param}[1]{{#1}} 

    \newcommand{\revcommon}[1]{{#1}}
    \newcommand{\reva}[1]{{#1}}
    \newcommand{\revb}[1]{{#1}}
    \newcommand{\revc}[1]{{#1}}
    
    \newcommand{\reve}[1]{{#1}}
    
    \newcommand{\revlabel}[2][1=]{}
    
    \newcommand{\dsnadd}[1]{{#1}}

    \usepackage[colorinlistoftodos,prependcaption,textsize=small]{todonotes}
    \paperwidth=\dimexpr \paperwidth + 4cm\relax
    \oddsidemargin=\dimexpr\oddsidemargin + 2cm\relax
    \evensidemargin=\dimexpr\evensidemargin + 2cm\relax
    \marginparwidth=\dimexpr \marginparwidth + 2cm\relax
    
    \newcommand{\atbcr}[2]{\ifnum#1=\value{version}\textcolor{red}{#2}\else{#2}\fi}
    \newcommand{\omcr}[2]{\ifnum#1=\value{version}\textcolor{blue}{#2}\else{#2}\fi}
    \newcommand{\atbcrcomment}[2]{\ifnum#1=\value{version}\todo[size=\scriptsize, linecolor=orange, bordercolor=orange, backgroundcolor=white]{\textcolor{red}{Atb:~#2}}\else{}\fi}
    \newcommand{\omcrcomment}[2]{\ifnum#1=\value{version}\todo[size=\scriptsize, linecolor=orange, bordercolor=orange, backgroundcolor=white]{\textcolor{blue}{Onur:~#2}}\else{}\fi}

\else
\ifshepherd

    \newcommand{\omcomment}[1]{}
    \newcommand{\omcommentinline}[1]{}

    \newcommand{\nrh}[0]{$HC_{first}$}

    \newcommand{\bscomment}[1]{}
    \newcommand{\bscommentinline}[1]{}

    \newcommand{\agy}[1]{{#1}}
    \newcommand{\agycomment}[1]{}
    
    \newcommand{\atb}[1]{{#1}}
    \newcommand{\atbcomment}[1]{}
    \newcommand{\atbcommentside}[1]{}
    \newcommand{\atatodo}[1]{}
    \newcommand{\atbnote}[1]{}

    \newcommand{\majdcomment}[1]{}
    \newcommand{\majdtodo}[1]{}

    \newcommand{\yctcomment}[1]{}

    \newcommand{\stecomment}[1]{}

    \newcommand{\param}[1]{{#1}} 

    \newcommand{\revcommon}[1]{{#1}}
    \newcommand{\reva}[1]{{#1}}
    \newcommand{\revb}[1]{{#1}}
    \newcommand{\revc}[1]{{#1}}
    
    \newcommand{\reve}[1]{{#1}}
    
    \newcommand{\revlabel}[2][1=]{}
    
    \newcommand{\dsnadd}[1]{{#1}}

    \newcommand{\atbcr}[2]{\ifnum#1=\value{version}\textcolor{red}{#2}\else{#2}\fi}
    \newcommand{\omcr}[2]{\ifnum#1=\value{version}\textcolor{red}{#2}\else{#2}\fi}
    \newcommand{\atbcrcomment}[2]{}
    \newcommand{\omcrcomment}[2]{}

\else

    \newcommand{\omcomment}[1]{}
    \newcommand{\omcommentinline}[1]{}

    \newcommand{\nrh}[0]{$HC_{first}$}

    \newcommand{\bscomment}[1]{}
    \newcommand{\bscommentinline}[1]{}

    \newcommand{\agy}[1]{{#1}}
    \newcommand{\agycomment}[1]{}
    
    \newcommand{\atb}[1]{{#1}}
    \newcommand{\atbcomment}[1]{}
    \newcommand{\atbcommentside}[1]{}
    \newcommand{\atatodo}[1]{}
    \newcommand{\atbnote}[1]{}

    \newcommand{\majdcomment}[1]{}
    \newcommand{\majdtodo}[1]{}

    \newcommand{\yctcomment}[1]{}

    \newcommand{\stecomment}[1]{}

    \newcommand{\param}[1]{{#1}} 

    \newcommand{\revcommon}[1]{{#1}}
    \newcommand{\reva}[1]{{#1}}
    \newcommand{\revb}[1]{{#1}}
    \newcommand{\revc}[1]{{#1}}
    
    \newcommand{\reve}[1]{{#1}}
    
    \newcommand{\revlabel}[2][1=]{}
    
    \newcommand{\dsnadd}[1]{{#1}}

    \newcommand{\atbcr}[2]{#2}
    \newcommand{\omcr}[2]{#2}
    \newcommand{\atbcrcomment}[2]{}
    \newcommand{\omcrcomment}[2]{}

    \usepackage[disable]{todonotes}
\fi
\fi
\fi
\fi
\fi

\newcommand{\ext}[1]{#1}

\def\UrlBreaks{\do\/\do-\/\do.\/\do:}

\expandafter\def\expandafter\UrlBreaks\expandafter{\UrlBreaks
  \do\a\do\b\do\c\do\d\do\e\do\f\do\g\do\h\do\i\do\j
  \do\k\do\l\do\m\do\n\do\o\do\p\do\q\do\r\do\s\do\t
  \do\u\do\v\do\w\do\x\do\y\do\z\do\A\do\B\do\C\do\D
  \do\E\do\F\do\G\do\H\do\I\do\J\do\K\do\L\do\M\do\N
  \do\O\do\P\do\Q\do\R\do\S\do\T\do\U\do\V\do\W\do\X
  \do\Y\do\Z}

\newcommand{\exploitingRowHammerAllCitations}[0]{\cite{fournaris2017exploiting, poddebniak2018attacking, tatar2018throwhammer, carre2018openssl, barenghi2018software, zhang2018triggering, bhattacharya2018advanced, google-project-zero, kim2014flipping, rowhammergithub, seaborn2015exploiting, van2016drammer, gruss2016rowhammer, razavi2016flip, pessl2016drama, xiao2016one, bosman2016dedup, bhattacharya2016curious, burleson2016invited, qiao2016new, brasser2017can, jang2017sgx, aga2017good, mutlu2017rowhammer, tatar2018defeating, gruss2018another, lipp2018nethammer, van2018guardion, frigo2018grand, cojocar2019eccploit,  ji2019pinpoint, mutlu2019rowhammer, hong2019terminal, kwong2020rambleed, frigo2020trrespass, cojocar2020rowhammer, weissman2020jackhammer, zhang2020pthammer, yao2020deephammer, deridder2021smash, hassan2021utrr, jattke2022blacksmith, tol2022toward, kogler2022half, orosa2022spyhammer, zhang2022implicit, liu2022generating, cohen2022hammerscope, zheng2022trojvit, fahr2022frodo, tobah2022spechammer, rakin2022deepsteal, aydin2022cyber, mus2022jolt, wang2022research, lefforge2023reverse,fahr2022effects, kaur2022work, cai2022feasibility, li2022cyberradar, roohi2022efficient, staudigl2022neurohammer, yang2022socially, islam2022signature}}

\newcommand{\mitigatingRowHammerAllCitations}[0]{\cite{AppleRefInc, rh-hp,rh-lenovo,greenfield2012throttling, kim2014flipping, kim2014architectural, bains14d, bains14c, aweke2016anvil, bains-merged, son2017making, seyedzadeh2018cbt,irazoqui2016mascat, you2019mrloc, lee2019twice, park2020graphene, yaglikci2021security, yaglikci2021blockhammer, frigo2020trrespass, kang2020cattwo, hassan2021utrr, qureshi2022hydra, saileshwar2022randomized, brasser2017can, konoth2018zebram, van2018guardion, vig2018rapid,  kim2022mithril, lee2021cryoguard, marazzi2022protrr, zhang2022softtrr, joardar2022learning, juffinger2023csi, yaglikci2022hira, saxena2022aqua, enomoto2022efficient, manzhosov2022revisiting, ajorpaz2022evax, naseredini2022alarm, joardar2022machine, hassan2022case, zhang2020leveraging,loughlin2021stop, devaux2021method, han2021surround, fakhrzadehgan2022safeguard, saroiu2022price, saroiu2022configure, loughlin2022moesiprime, zhou2022lt, hong2023dsac, mutlu2023fundamentally, marazzi2023rega, di2023copy, sharma2022review, woo2023scalable, park2022row, wi2023shadow, kim2023ddr5, gude2023defending, guha2022criticality, france2022modeling, france2022reducing, bennett2021panopticon, arikan2022processor, tomita2022extracting, saxena2023pt, zhou2023dnndefender, woo2023rampart, kim2023how}}

\newcommand{\rowHammerGetsWorseCitations}[0]{\cite{kim2020revisiting, frigo2020trrespass, yaglikci2022understanding, orosa2021deeper, mutlu2017rowhammer, mutlu2018rowhammer, mutlu2019rowhammer, mutlu2023fundamentally}}

\makeatletter
\g@addto@macro{\normalsize}{%
  \setlength{\abovedisplayskip}{3pt plus 0.5pt minus 1pt}
  \setlength{\belowdisplayskip}{3pt plus 0.5pt minus 1pt}
  \setlength{\abovedisplayshortskip}{0pt}
  \setlength{\belowdisplayshortskip}{0pt}
  \setlength{\intextsep}{4pt plus 1pt minus 1pt}
  \setlength{\textfloatsep}{4pt plus 1pt minus 1pt}
  \setlength{\skip\footins}{5pt plus 1pt minus 1pt}}
\makeatother

\usepackage{titlesec}
\titlespacing\section{0pt}{2pt plus 1pt minus 1pt}{3pt plus 1pt minus 2pt}
\titlespacing\subsection{0pt}{2pt plus 1pt minus 1pt}{3pt plus 1pt minus 2pt}
\titlespacing\subsubsection{0pt}{2pt plus 1pt minus 1pt}{3pt plus 1pt minus 2pt}

\newcounter{obs}
\newcommand\observation[1]{%
   \refstepcounter{obs}
   \noindent
   \colorbox{gray!20}{\textbf{\obslbl{} \theobs.}} \emph{#1}}
   
\newcounter{take}
\newcommand\take[1]{%
   \refstepcounter{take}
   \vspace{1mm}
  \noindent
  \begin{tabular}{|p{0.95\linewidth}|}
       \hline
       \textbf{{Takeaway \thetake}.} {{#1}}\\
       \hline 
  \end{tabular}
  
}

\newcommand{\figref}[1]{Fig.~\ref{#1}}

\newcommand{\secref}[1]{\S~\ref{#1}}

\newcommand{\obslbl}[0]{Obsv.}

\usepackage{glossaries}

\newacronym{vdd}{$V_{DD}$}{supply voltage}
\newacronym{vpp}{$V_{PP}$}{wordline voltage}
\newacronym{vppmin}{$V_{PPmin}$}{the lowest \gls{vpp} at which the DRAM module can successfully communicate with the FPGA}
\newacronym{vwl}{$V_{PP}$}{wordline voltage}
\newacronym{gnd}{$GND$}{ground}
\newacronym{hcfirst}{$HC_{first}$}{the minimum \agy{hammer} count necessary to cause a RowHammer bitflip}
\newacronym{taggon}{$t_{AggON}$}{the aggressor row on time}
\newacronym{rblast}{$r_{Blast}$}{blast radius}
\newacronym{ber}{$BER$}{the fraction of DRAM cells that experience a bitflip in a DRAM row}
\newacronym{nhc}{$N_{HC}$}{hammer count}
\newacronym{hc}{$HC$}{hammer count}
\newacronym{trcd}{$t_{RCD}$}{row activation latency}
\newacronym{tcl}{$t_{CL}$}{column access latency}
\newacronym{tcwl}{$t_{CWL}$}{column write latency}
\newacronym{trp}{$t_{RP}$}{precharge latency}
\newacronym{trcdmin}{$t_{RCDmin}$}{{the minimum time delay required}}
\newacronym{tras}{$t_{RAS}$}{charge restoration latency}
\newacronym{trasmin}{$t_{RASmin}$}{the minimum latency required}
\newacronym{trefw}{$t_{REFW}$}{refresh window}
\newacronym{trefi}{$t_{REFI}$}{refresh interval}
\newacronym{vgs}{$V_{GS}$}{gate-to-source voltage}
\newacronym{vthresh}{$V_{TH}$}{the voltage threshold that the bitline voltage should exceed for the activation to be reliably completed}
\newacronym{kde}{KDE}{kernel density estimate}
\newacronym{ref}{refresh}{$REF$}
\newacronym{tsv}{through-silicon-via}{TSV}

\glsdisablehyper

\newcounter{version}
\ifiterations
\else
\fi

\usepackage[available,functional]{usenixbadges}

\begin{document}

\bstctlcite{IEEEexample:BSTcontrol}
\title{\LARGE{Read Disturbance in High Bandwidth Memory:\\A Detailed Experimental Study on HBM2 DRAM Chips}}
\setlength{\droptitle}{-2em}

\author{\vspace{-18pt}\\%
\fontsize{11}{12}\selectfont%
{Ataberk Olgun$^1$}\quad%
{Majd Osseiran$^1$}\quad%
{A. Giray Ya\u{g}l{\i}k\c{c}{\i}$^1$}\quad%
{Yahya Can Tu\u{g}rul$^1$}\quad%
\\%
\fontsize{11}{12}\selectfont
{Haocong Luo$^1$}\quad
{Steve Rhyner$^1$}\quad
{Behzad Salami$^2$}\quad
{Juan Gomez Luna$^1$}\quad%
{Onur Mutlu$^1$}\quad
\vspace{0pt}\\%
{\fontsize{10}{11}\selectfont
$^1$\emph{ETH Z{\"u}rich}
\qquad
$^2$\emph{BSC}%
}
}

\maketitle

\ifiterations
    \fancyhead{}
    \fancyhead[C]{\textcolor{blue}{\emph{Version 1.2~---~\today, \ampmtime}}}
    \fancypagestyle{firstpage}
    {
        \fancyhead{}
        \fancyhead[C]{\textcolor{red}{CONFIDENTIAL DRAFT -- DO NOT DISTRIBUTE -- TO APPEAR IN DSN'24} \\ \textcolor{blue}{\emph{Version 1.0~---~\today, \ampmtime}} }
    }
    \pagenumbering{arabic}
    \thispagestyle{fancy}
    \pagestyle{plain}
\else 
    \pagenumbering{arabic}
    \renewcommand{\headrulewidth}{0pt}
    \fancyhf{} 
    \fancyfoot{} 
    \thispagestyle{plain}
    \pagestyle{plain}
\fi

\begin{abstract}

We experimentally demonstrate the effects of read disturbance (RowHammer and RowPress) and
uncover the inner workings of undocumented read disturbance defense 
mechanisms in High Bandwidth Memory (HBM). Detailed characterization of six 
real HBM2 DRAM chips \revlabel{Rev.B-C2}\revb{in two different FPGA boards} shows that 
(1) the read disturbance vulnerability 
significantly varies between different HBM2 chips and between
different components (e.g., 3D-stacked channels) inside a chip,
(2) DRAM rows at the end and in the middle of a bank are more resilient to read disturbance,
(3) fewer additional activations are sufficient to induce more read disturbance bitflips 
in a DRAM row if the row exhibits the first bitflip at a relatively high 
activation count, 
(4) a modern HBM2 chip implements undocumented read disturbance defenses that track potential aggressor rows based on how many times they are activated. 
We describe how 
our findings could be leveraged to develop more powerful read disturbance 
attacks and more efficient defense mechanisms.
\atbcr{1}{We open source all our code and data to facilitate future research} \omcr{2}{at \url{https://github.com/CMU-SAFARI/HBM-Read-Disturbance}}.

\end{abstract}

\section{Introduction}
\label{sec:introduction}

{Modern DRAM chips} {{suffer from} read disturbance issues~\cite{kim2014flipping,mutlu2019retrospective,mutlu2023fundamentally,luo2023rowpress} that can be exploited to break memory isolation, threatening the \omcr{2}{robustness (including} safety, security, and reliability\omcr{2}{)} of modern DRAM-based computing systems. RowHammer~\cite{kim2014flipping} and RowPress~\cite{luo2023rowpress} are two prominent examples of read disturbance. Repeatedly opening/activating and closing a DRAM row (i.e., aggressor row) \emph{many times} (e.g., thousands \omcr{2}{of times}) induces \emph{RowHammer bitflips} in physically nearby rows (i.e., victim rows). Keeping the aggressor row open for a long period of time (i.e., a large aggressor row on time, $t_{AggON}$) amplifies the effects of read disturbance and induces \emph{RowPress bitflips}, \omcr{2}{at much lower aggressor row activation counts}~\cite{luo2023rowpress}.
{N}umerous studies {demonstrate that} a malicious attacker can {reliably} cause {{read disturbance} bitflips} in a targeted manner to compromise system integrity, confidentiality, and availability~\exploitingRowHammerAllCitations{}. 
{{Read disturbance worsens} {in new DRAM chips with {smaller technology nodes}, {where}} {RowHammer} bitflips 1)~{happen} with fewer row activations, e.g., $10\times$ reduction in less than a decade~\cite{kim2020revisiting} and 2)~{appear} {in} more DRAM cells, compared to old DRAM {chips}~\rowHammerGetsWorseCitations{}.}

{To meet the high-bandwidth requirements of modern data-intensive applications {(e.g., GPU workloads~\cite{bakhoda2009analyzing,che2009rodinia}, \omcr{2}{machine learning training and inference models~\cite{jouppi2023tpuv4,brown2020language,devlin2019bert}})}, DRAM designers develop High Bandwidth Memory (HB{M})~\cite{jedec2021hbm} DRAM chips, which contain multiple layers of 3D-stacked DRAM dies, using cutting-edge technology nodes.}\footnote{{We use ``chip'' to refer to an \emph{HBM2 stack}. An HBM2 stack contains one or multiple DRAM layers. We refer to each such layer using ``DRAM die''.}}
{It is important to understand {read disturbance} in HBM DRAM chips \atbcr{2}{because 1)} they have new architectural characteristics (e.g., multiple layers of DRAM dies, area- and energy-intensive through-silicon vias), which might affect the chip's read disturbance vulnerability in currently-unknown ways,}
\atbcr{2}{and 2) they are extensively used in \omcr{3}{critical system} infrastructures of today (\omcr{3}{e.g., machine learning} training and inference~\cite{jouppi2023tpuv4,nvidiaA100,nvidiaH100,jouppi2020domain,amd-cdna,amd-cdna2})\omcrcomment{3}{these references seem off}.}
{Such understanding can help identify potential {read-disturbance}-induced security and reliability issues in HBM-based systems and allow for effective and efficient defense mechanisms.}

{\textbf{Our goal} is to experimentally {analyze} how vulnerable \dsnadd{HBM DRAM \atbcr{1}{chips are}} to read disturbance.}
To this end, we provide {the first detailed} experimental characterization {of the {RowHammer} and the RowPress vulnerabilit\omcr{2}{ies}} in six {modern} HBM2 DRAM chips \revlabel{Rev.B-C2}\revb{in two different FPGA boards}. We {provide} \param{four} main analyses in our study. 
{First, we analyze the spatial variation in RowHammer vulnerability ({\secref{sec:spatial-variation-analysis}) based on the physical location of victim rows in terms of two metrics: \gls{ber} and \gls{hcfirst}.}
\revlabel{Rev.B-C1}\revb{We use these two metrics to quantify the RowHammer susceptibility of a DRAM row. For example, a row with a small \gls{ber} value is susceptible to only a \omcr{2}{small number of} RowHammer-induced bitflips. \omcr{2}{As such,} this row is more \emph{RowHammer-resilient} than other DRAM rows with higher \gls{ber} values.}
Second, we analyze the number of aggressor row activations (i.e., hammer count) necessary to induce the first 10 bitflips in a DRAM row (\secref{sec:hcnth}). 
We demonstrate how many additional \omcr{2}{hammers} over \gls{hcfirst} \omcr{2}{are} needed to induce each of the first 10 bitflips.
{Third, we test RowPress and RowHammer's sensitivities to the amount of time a row remains active{, i.e.,} \gls{taggon} (\secref{sec:taggon}). To do so, we sweep {\gls{taggon}} from the minimum standard \omcr{2}{value of} \SI{29.0}{\nano\second}\atbcrcomment{2}{The HBM2 standard does not report this number. We cannot find manufacturer-specific datasheets. The number is from Mike O'Connor's thesis and published work.}\atbcrcomment{3}{We cite the proper references where we describe our methodology} to an extreme \SI{16.0}{\milli\second}.} 
{Fourth}, we investigate undocumented in-DRAM {read disturbance} defense mechanisms {that are triggered by periodic refresh operations} \revlabel{Rev.B-C1}\revb{(e.g., Target Row Refresh~\cite{hassan2021utrr,micron2018ddr4trr,frigo2020trrespass}, or TRR for short)} in an HBM2 chip {(\secref{sec:uncovering})}.\footnote{{\label{foot:trr}The HBM2 standard specifies a Target Row Refresh (TRR) Mode. To enable TRR Mode, the memory controller issues a well-defined series of commands. {We investigate undocumented TRR mechanisms that function even when the DRAM chip is \emph{not} in TRR mode.}
}} {We summarize the key observations from our four main analyses.}

\noindent
{\textbf{1)~Spatial variation in RowHammer~(\secref{sec:spatial-variation-analysis}).}} 
{{First}, DRAM rows near the end and in the middle of a DRAM bank (last and middle \param{832} rows) exhibit substantially smaller \gls{ber} than other DRAM rows.}
{{Second}}, RowHammer \gls{ber} and \gls{hcfirst} vary between DRAM chips. For example, the chip-level mean \gls{ber} and minimum \gls{hcfirst} differ by up to {\param{0.49} \revb{percentage points (pp)}\revlabel{Rev.B-C1} and \param{3556}, respectively. {{Third}}, different 3D-stacked channels of an HBM2 chip exhibit significantly different levels of RowHammer vulnerability in \gls{ber} (\revcommon{e.g., the channel with the highest mean \gls{ber} has 1.99$\times{}$ the mean \gls{ber} of the channel with the lowest mean \gls{ber}}) and \gls{hcfirst}. {Fourth, the mean \gls{ber} variation across channels in multiple HBM2 chips is larger than the mean \gls{ber} variation across all HBM2 chips, for all tested data patterns (Table~\ref{table_data_patterns}).\atbcrcomment{3}{This (For example,...) is complicated to describe here. Can I remove it?  Not very important to highlight}
\atbcr{2}{For example, the channel with the highest mean \gls{ber} across all its rows in Chip \param{4} has \param{0.88} \revb{pp} higher \gls{ber} than the channel with that of smallest and the chip with the highest \gls{ber} across all its rows has \param{0.38} \revb{pp} higher \gls{ber} than the chip with that of smallest.}}



\noindent
\textbf{2)~RowHammer's Sensitivity to Hammer Count~(\secref{sec:hcnth}).} We show that the hammer count to induce more than one \omcr{2}{or more additional} RowHammer bitflips (up to 10) in a row can be very close to (\revb{\param{1.15}$\times{}$} larger than) or very far from (\param{5.22}$\times{}$ larger than) the hammer count to induce the \emph{first} RowHammer bitflip depending on the DRAM row. We find that, in general, it takes fewer additional hammer counts (over \nrh{}) to induce up to 10 RowHammer bitflips in a DRAM row that has a large \nrh{} compared to a DRAM row that has a small \nrh{}.

\noindent
\textbf{3)~RowHammer's and RowPress's Sensitivities to \gls{taggon}~(\secref{sec:taggon}).} {We observe that as the time an aggressor row remains open {(\gls{taggon})} increases, DRAM cells become more vulnerable to read disturbance. \revcommon{For example, the \gls{hcfirst} of a row at a \gls{taggon} of \SI{35.1}{\micro\second} \atbcr{2}{(the maximum allowed time to keep a row open~\cite{jedec2021hbm})} is} \param{222.57}$\times{}$ \revcommon{smaller than the \gls{hcfirst} of the row at a \gls{taggon} of \SI{29.0}{\nano\second}, averaged across all tested DRAM rows.}
\revcommon{We observe that \emph{only} one DRAM row activation is sufficient to induce RowPress bitflips at an} extreme \gls{taggon} of \SI{16}{\milli\second}.}

\noindent
{\textbf{4)~In-DRAM RowHammer defenses~(\secref{sec:uncovering}).} We uncover that an HBM2 DRAM chip implements an in-DRAM RowHammer defense mechanism that is not disclosed in the HBM2 specification~\cite{jedec2021hbm}. {The undocumented target row refresh (TRR) mechanism \omcr{2}{identifies as \atbcr{4}{aggressor} rows} i) the \emph{first row} that gets activated after a TRR operation (a victim row refresh) and ii) the row whose activation count exceeds half the number of total row activations within a refresh interval.}}
We experimentally demonstrate that \dsnadd{an} attacker \dsnadd{could} practically and reliably defeat this undocumented TRR mechanism in real HBM2 DRAM chips by leveraging our observations.

We highlight \param{three} of the \param{six} key implications of our observations for future read disturbance attacks and defenses (\secref{sec:implications}): 1)~the maximum \gls{ber} (247 bitflips in a row of 8192 bits) is likely sufficient for conducting practical attacks (e.g., privilege escalation) \atbcr{2}{in} modern HBM-based systems, 2)~a read disturbance attack could find exploitable bitflips faster by targeting the most-read-disturbance-vulnerable HBM2 channels, and 3)~read disturbance defense mechanisms could \omcr{2}{more efficiently prevent \atbcr{3}{read-disturb} bitflips by} adapt\omcr{2}{ing} to the heterogeneous distribution of the RowHammer and RowPress vulnerability across channels and subarrays. 




{We make the following contributions:}
\begin{itemize}
    \item We present the first detailed experimental characterization of read disturbance (RowHammer and RowPress) in {six} state-of-the-art HBM2 DRAM chip{s}. We show that {all of the tested} HBM2 DRAM chips are susceptible to read disturbance bitflips.
    \item We show that the RowHammer and RowPress vulnerability in HBM2 varies significantly across {chips} and HBM2 components within each chip (e.g., 3D-stacked channels, pseudo channels, banks, and rows).
    \item We present the first analysis on hammer count to induce up to 10 bitflips in a DRAM row. We show that a DRAM row with a large hammer count to induce the first bitflip is likely to require fewer additional hammer counts to exhibit the next 9 bitflips. 
    \item We characterize the RowPress vulnerability in six HBM2 chips. \atbcr{2}{Keeping the aggressor row open for the maximum allowed time reduces the activation count to induce a bitflip by three orders of magnitude on average across all tested chips.} We show that all chips exhibit bitflips \atbcr{2}{when the aggressor row is activated \emph{only once} and} kept open for a very long time (\SI{16}{\milli\second}). 
    \item We uncover the inner workings of an undocumented in-HBM-chip read disturbance defense mechanism. We {analyze this defense mechanism and} craft a specialized {RowHammer access} pattern that bypasses the defense.
    \item \atbcr{1}{We open-source all our infrastructure~\cite{self.github}, test programs, and raw data to enable 1) reproduction and replication of our results, and 2) further research on read disturbance in HBM chips.}
\end{itemize}
\section{Background \& Motivation}

We describe the necessary background on HBM2 organization and operation\omcr{3}{, and motivate read disturbance and its implications for real HBM2 systems.}

\subsection{HBM2 Organization}

{\figref{fig:hbm-organization} shows the organization of an HBM2 DRAM chip~\cite{oconnor2021thesis} used in an FPGA-based system. The memory controller communicates with multiple stacks of HBM using the HBM2 interface (\dingOne{}). An HBM2 stack contains multiple DRAM dies stacked on top of the buffer die and connected using through-silicon vias (TSVs) (\dingTwo{}). Each HBM2 die comprises one or multiple HBM2 channels that can operate independently (\dingThree{}). An HBM2 channel contains multiple pseudo channels and each pseudo channel has multiple banks (\dingFour{}). A DRAM bank comprises multiple DRAM cells that are laid in a two-dimensional array of rows and columns (\dingFive{}). {DRAM cells are typically partitioned into multiple DRAM subarrays~\cite{kim2012case,seshadri2013rowclone,chang2014improving} (not shown in the figure) that each contain a row buffer.} When enabled, a wordline connects a DRAM cell to its bitline, copying the data stored in the DRAM cell to the row buffer.}
\begin{figure}[!h]
    \centering
    \includegraphics[width=\linewidth]{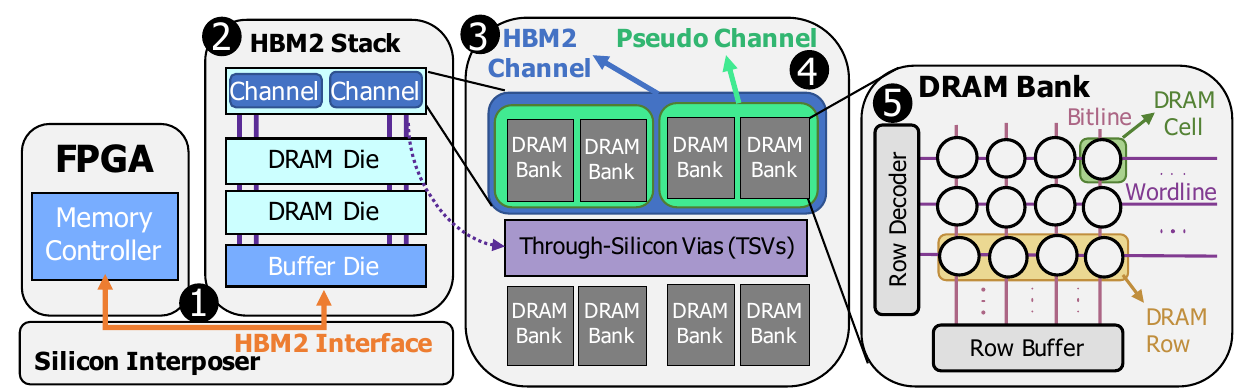}
    \caption{HBM2 DRAM system organization}
    \label{fig:hbm-organization}
\end{figure}

\subsection{\atb{HBM2 Operation}}

{To access a DRAM chip, the memory controller needs to issue the following sequence of commands. First, the controller issues an activate (ACT) command targeting a DRAM row to access a DRAM cell. The row decoder enables the row's wordline, copying the data in the row to the row buffer. Second, to read from or write to a particular column in that row, a RD or WR command needs to be issued. Finally, when accesses to the open row are complete, the memory controller issues a precharge ($PRE$) command, which disables the enabled wordline so that the memory controller can later access a different cell in another DRAM row.}

{To maintain reliable DRAM operation, the memory controller must obey manufacturer-recommended, standard timing parameters. These timing parameters ensure that the DRAM circuitry has enough time to execute the operations dictated by DRAM commands. Two relevant timing parameters for our study are \gls{tras} and \gls{trefi}. First, \gls{tras}, the minimum time that a row should remain active before a $PRE$ command is sent to the row's bank. $t_{RAS}$ guarantees that DRAM sense amplifiers have enough time to restore charge in cells of the open row before the row is closed. Second, \omcr{3}{\gls{trefi}}, the periodic interval at which a refresh cycle is required. Since a DRAM cell stores data as charge in its capacitor and the capacitor naturally loses charge over time, the capacitor must be periodically refreshed to prevent data corruption. Consequently, the memory controller should issue a $REF$ command on average every \SI{3.9}{\micro\second} \cite{jedec2021hbm}} ($t_{REFI}$), such that every DRAM cell is refreshed once at a fixed refresh window (e.g., \SI{32}{\milli\second}). The memory controller may delay a $REF$ command up to \SI{35.1}{\micro\second} ($9*t_{REFI}$). However, any such large delay must be compensated by multiple smaller delays between successive $REF$ commands following the large delay.

\noindent



\subsection{{Motivation}}

Read-disturb phenomena (e.g., RowHammer~\cite{kim2014flipping} and RowPress~\cite{luo2023rowpress}) break the fundamental building block of modern system security principles, i.e., \emph{memory isolation}.
This property allows read-disturb phenomena to be used in system-level attacks that compromise system integrity, confidentiality, and availability in various real computing systems, as many prior works have shown~\exploitingRowHammerAllCitations. 
Therefore, it is critical to understand the properties of the RowHammer and RowPress vulnerabilit\omcr{3}{ies} to design defense mechanisms and protect modern DRAM chips against read-disturbance-based attacks. 
Unfortunately, no prior work extensively studies the RowHammer and RowPress vulnerabilit\omcr{3}{ies} of modern HBM chips. 
To this end, our goal is to \omcr{3}{experimentally} evaluate and understand the RowHammer and RowPress vulnerabilit\omcr{3}{ies} in real HBM chips. To achieve this goal, we perform a rigorous experimental characterization study of read disturbance on six HBM2 chips in two different types of integrated circuit packages.

\section{Experimental Infrastructure}

We \atbcr{1}{use} a modified version of the DRAM Bender testing infrastructure~\cite{olgun2023drambender,safari-drambender}. This infrastructure allows us to precisely control the HBM2 command timings at the granularity of \revb{\SI{1.67}{\nano\second}} (i.e., the HBM2 interface clock speed is \param{600} MHz). 
{{All tested} HBM2 chip{s} ha{ve} i) 8 channels, ii) 2 pseudo channels, iii) 16 banks, iv) 16384 rows, and v) 1 KiB rows.}


\noindent
\textbf{Testing setup.}
\figref{fig:dram-bender} shows {one of} our {six} testing setup{s}. We conduct experiments using one Bittware XUPVVH~\cite{xupvvh} (1) {and five AMD Xilinx Alveo U50 FPGA boards~\cite{u50} (not shown in~\figref{fig:dram-bender})}.
We use the heating pad (2) and the cooling fan (3) to increase and reduce the temperature of the HBM2 chip, respectively. The Arduino~\cite{arduinomega} temperature controller (4) communicates with i)~the host machine to retrieve a target temperature and ii)~the FPGA board to retrieve the HBM2 chip's temperature. A host machine executes the test programs described in \secref{sec:testing-methodology} on the FPGA board using the PCIe connection (5).



\begin{figure}[h]
    \centering
    \includegraphics[width=1.0\linewidth]{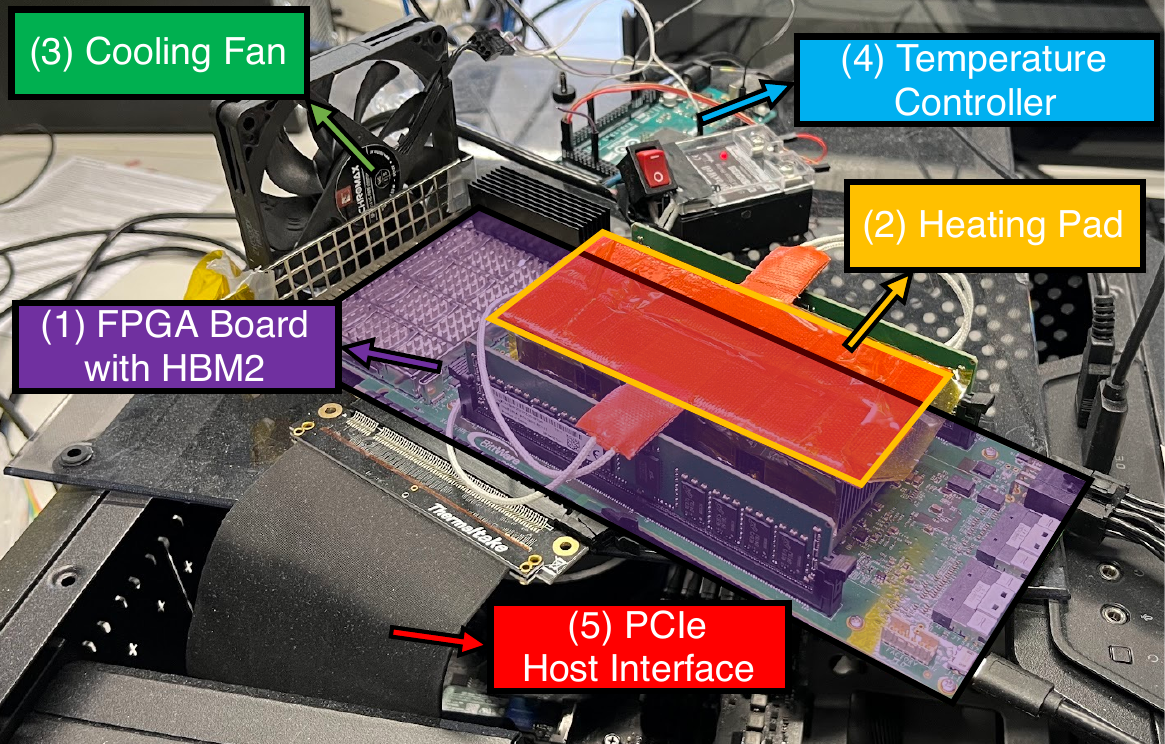}
    \caption{FPGA-based HBM2 DRAM tester.}
    \label{fig:dram-bender}
\end{figure}

\begin{figure*}[!th]
    \centering
    \includegraphics[width=\linewidth]{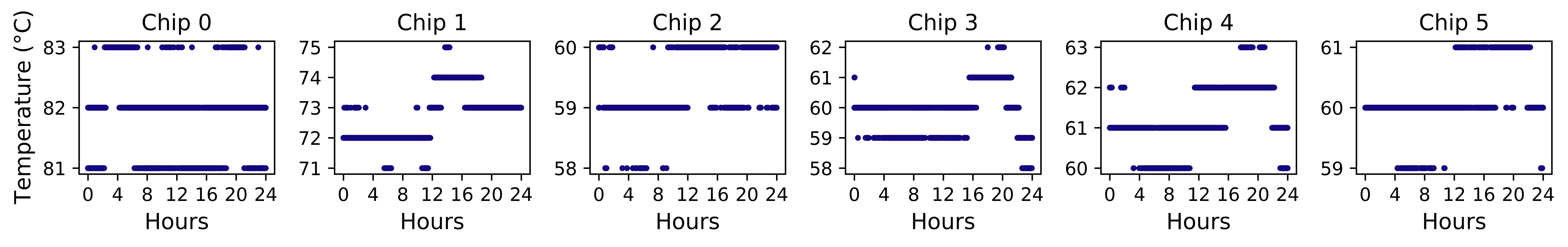}
    \vspace{-2em}
    \caption{Tested HBM2 chips' temperature over time (24 hours). We draw temperature measurements taken every 5 seconds over a 24 hour time window.}
    \vspace{-0.5em}
    \label{fig:temperature-measurements}
\end{figure*}

\subsection{Testing Methodology}
\label{sec:testing-methodology}

\noindent
\textbf{Disabling Sources of Interference}. 
{We identify \param{four} sources that can interfere with our characterization results: 1) periodic refresh~\cite{jedec2021hbm}, 2) on-die read disturbance defense mechanisms (e.g., TRR~\cite{frigo2020trrespass,hassan2021utrr,micron2018ddr4trr}), 3) data retention failures~\cite{liu2013experimental,patel2017reaper}, and 4) ECC~\cite{jedec2021hbm}. First, we do \emph{not} issue periodic refresh commands in our experiments. Second, disabling periodic refresh disables all known on-die read disturbance defense mechanisms~\cite{orosa2021deeper,yaglikci2022understanding,kim2020revisiting,hassan2021utrr}. Third, we ensure that our experiments finish within the \SI{32}{\milli\second} refresh interval where manufacturers guarantee no retention errors will occur~\cite{jedec2021hbm}. Fourth, we disable ECC \atbcr{3}{by setting the corresponding HBM2 mode register bit to zero~\cite{jedec2021hbm}}.}

\noindent
\textbf{RowHammer and RowPress Access Pattern}. 
We use the double-sided read disturbance access pattern~\cite{kim2014flipping,kim2020revisiting,orosa2021deeper,seaborn2015exploiting}, which alternately activates each aggressor row. 
{We record the bitflips observed \omcr{3}{in} the sandwiched victim row (i.e., the row between two aggressor rows).}

\noindent \textbf{Logical-to-Physical Row Mapping}. {DRAM manufacturers use mapping schemes to translate logical (memory-controller-visible) addresses to physical row addresses~\cite{kim2014flipping, smith1981laser, horiguchi1997redundancy, keeth2001dram, itoh2013vlsi, liu2013experimental,seshadri2015gather, khan2016parbor, khan2017detecting, lee2017design, tatar2018defeating, barenghi2018software, cojocar2020rowhammer,  patel2020beer, yaglikci2021blockhammer, orosa2021deeper}.}
To identify aggressor rows that are physically adjacent to a victim row, we reverse-engineer the row mapping scheme {following the methodology described in prior work~\cite{orosa2021deeper}}. 


\noindent
\textbf{RowHammer and RowPress Test Parameters}.
{We configure our tests by tuning \param{\omcr{3}{three}} parameters: 
1)~Hammer count: We define the \emph{hammer count} of a double-sided read disturbance access pattern as the number of activations \emph{\omcr{3}{each}} aggressor row receives. Therefore, during a double-sided RowHammer or a RowPress test with a hammer count of 10, we activate each of the two aggressor rows 10 times, resulting in a total of 20 row activations.
2)~\Glsfirst{taggon}: The time each aggressor row stays on with each activation during a RowHammer or a RowPress test.}
3)~Data pattern: We use the four data patterns {(Table~\ref{table_data_patterns}) that are widely used in memory reliability testing~\cite{vandegoor2002address} and by prior work on DRAM characterization (e.g.,~\cite{kim2014flipping,kim2020revisiting,orosa2021deeper,luo2023rowpress,yaglikci2024svard})}. \revlabel{Rev.B-C1}\revb{{For each DRAM row,} we define \emph{WCDP} as the data pattern that causes the smallest \gls{hcfirst}. {When multiple} data patterns cause the same \gls{hcfirst}, we select WCDP as the data pattern that causes the largest \gls{ber} at hammer count = 256K.}

\begin{table}[htbp]
\caption{Data patterns used in our experiments}
\vspace{-1.5em}
\begin{center}
\begin{adjustbox}{max width=\linewidth}
\begin{tabular}{|c||c|c|c|c|}
\hline
\textbf{Row Addresses} & \textbf{\textit{Rowstripe0}}& \textbf{\textit{Rowstripe1}}& \textbf{\textit{Checkered0}} & \textbf{\textit{Checkered1}}\\
\hline
\hline
Victim (V) & 0x00 & 0xFF & 0x55 & 0xAA\\
\hline
Aggressors (V $\pm$ 1) & 0xFF & 0x00 & 0xAA & 0x55\\
\hline
V $\pm$ [2:8] & 0x00 & 0xFF & 0x55 & 0xAA\\
\hline
\end{tabular}
\end{adjustbox}
\vspace{-1em}
\label{table_data_patterns}
\end{center}
\end{table}

\noindent
\textbf{\atbcr{3}{Read Disturbance} Vulnerability Metrics}.
{We measure \atbcr{3}{the read disturbance} vulnerability based on two metrics: 1)~\glsfirst{hcfirst} and 2)~\glsfirst{ber} as defined in \secref{sec:introduction}.}


\noindent
\textbf{Tested DRAM Components}.
{To {maintain a reasonable experiment time}, the number of DRAM components (channels, pseudo channels, banks, and rows) we test \atbcr{4}{varies} depending on the experiment type. Table~\ref{table_exp_param} summarizes the number of components tested for each experiment type.}

\begin{table}[htbp]
\caption{Tested DRAM components for each experiment type}
\vspace{-1.25em}
\begin{center}
\begin{adjustbox}{max width=\linewidth}
\begin{tabular}{|c||c|c|c|c|}
\cline{1-5}
\textbf{Experiment Type} & \textbf{{Rows (Per Bank)}}& \textbf{{Banks}}& \textbf{{Pseudo Channels}} & \textbf{{Channels}}\\
\hline
\hline
RowHammer \gls{ber} & 16384 & 1 & 1 & 8\\
\hline
RowHammer \gls{hcfirst} & 3072 & 3 & 2 & 8\\
\hline
RowPress \gls{ber} & 384 & 1 & 1 & 3\\
\hline
RowPress \gls{hcfirst} & 384 & 1 & 1 & 3\\
\hline
\end{tabular}
\end{adjustbox}
\vspace{-1em}
\label{table_exp_param}
\end{center}
\end{table}

\noindent
\revlabel{Rev.B-Q1}\revb{\textbf{Experiment Repetitions.} We repeat every experiment five times. We report 1) the average value across five repetitions for \gls{ber} experiments and 2) the minimum value across five repetitions for \gls{hcfirst} experiments.}

\noindent
\textbf{HBM2 Chip Labeling.} We \dsnadd{label the XUPVVH board's HBM2 chip as Chip 0 and the five U50 boards' HBM2 chips as Chip 1, 2, 3, 4, and 5.} 
%

\noindent
\revlabel{Rev.C-Q1}\revc{\textbf{Estimated ages of the tested HBM2 chips.} We do \emph{not} know the precise model information details (including the manufacturing date) of the HBM2 chips we test, as the manufacturer does not disclose the specifications for the HBM2 chips in their products~\cite{xilinxForumPost} and as our custom memory controller does not yet support accessing this information from the mode status registers of the HBM2 chip. We estimate the ages of the tested HBM2 chips based on the dates that we acquired them: Chip0 was 2 years and 9 months, Chip1 was 8 months, and Chip2-5 were 3 months old when we started our experiments.}


\noindent
\textbf{Temperature Control.} We use the temperature controller setup shown in \figref{fig:dram-bender} for Chip 0 and set the target temperature for this chip to \SI{82}{\celsius}.\atbcrcomment{3}{only chip 0 is set to 82c} \figref{fig:temperature-measurements} shows how the temperature of each chip varies during 24 hours based on measurements taken every 5 seconds. Even though we do not have the same temperature controller setups for the \omcr{3}{six} chips, we observe that their temperature is stable.\revlabel{Rev.B-Q3}\footnote{\revb{We retrieve the temperature for these five chips from an in-HBM2-chip temperature sensor using the IEEE 1500 test port~\cite{jedec2021hbm}.}} 



\section{Spatial Variation in RowHammer}
\label{sec:spatial-variation-analysis}

We provide the first detailed spatial variation analysis of RowHammer across channels, pseudo channels, banks, and rows in six HBM2 chips.

\subsection{\atb{RowHammer Across Chips}}
{\figref{fig:chipberplot} shows the distribution of \gls{ber} (y-axis) across all tested DRAM rows for each tested data pattern (x-axis) in all tested chips (color-coded). A higher \gls{ber} indicates worse RowHammer vulnerability as more DRAM cells in a row exhibit RowHammer bitflips.}

\begin{figure}[!h]
    \centering
    \includegraphics[width=\linewidth]{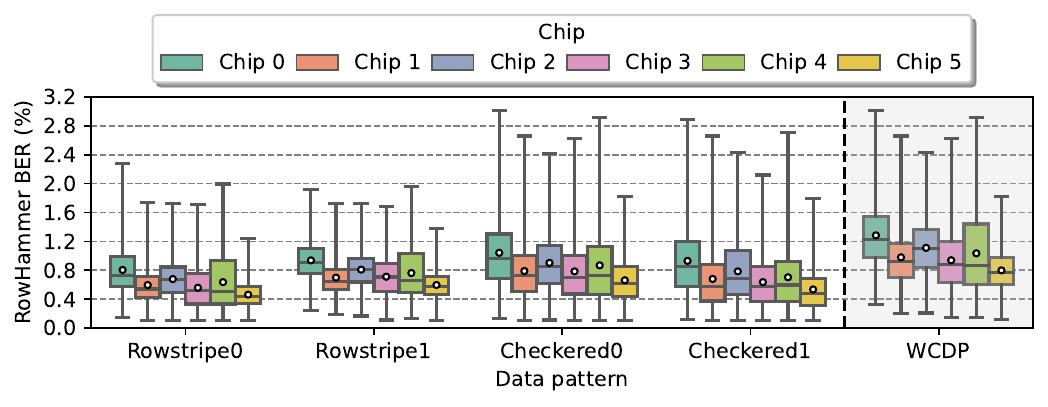}
    \caption{{\Glsfirst{ber} across different HBM2 chips. Error bars show the range of \gls{ber} across all tested rows in each chip.}}
    \label{fig:chipberplot}
\end{figure}

\observation{There are RowHammer bitflips in all tested DRAM rows in all chips. RowHammer \gls{ber} varies across chips.}



\revlabel{Rev.B-C1}For example, \revb{the \gls{ber} for a DRAM row in Chip 0 can reach up to 3.02\% (i.e., 3.02\% of all cells in a DRAM row exhibit RowHammer bitflips) and the mean \gls{ber} across all DRAM rows in Chip 0 is 1.04\%. In Chip 5, the highest \gls{ber} for a DRAM row is 1.82\% and the mean \gls{ber} across all DRAM rows is 0.66\%.}
We observe the most significant difference in \revb{mean} \gls{ber} \revb{across all rows in a chip} between Chip 0 (\param{1.28\%}) and Chip 5 (\param{0.80\%}) as {\param{0.49} \revb{percentage points (pp)}} for \revb{the worst case data pattern (WCDP, see~\secref{sec:testing-methodology})}.

\observation{Data pattern affects RowHammer \gls{ber}.}

As an example, the Checkered0 and Checkered1 data patterns result in substantially higher mean \gls{ber} across rows for every DRAM chip compared to the Rowstripe0 and Rowstripe1 data patterns. The mean \gls{ber} across all tested DRAM rows \revcommon{(across all chips)} is \param{0.76\%} and \param{0.67\%} for Checkered0/1 and Rowstripe0/1 data patterns, respectively.

\figref{fig:chiphcfplot} shows the distribution of \gls{hcfirst} (y-axis) across all tested DRAM rows for each tested data pattern (x-axis). A lower \gls{hcfirst} indicates worse RowHammer vulnerability as DRAM cells exhibit RowHammer bitflips with fewer aggressor row activations (i.e., it takes a shorter time to induce the first RowHammer bitflip in a row).

\begin{figure}[!h]
    \centering
    \includegraphics[width=\linewidth]{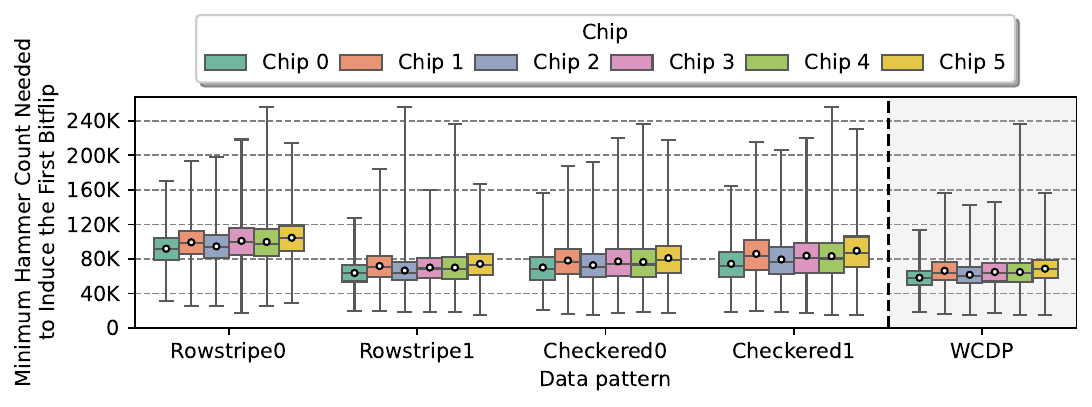}
    \caption{{\Glsfirst{hcfirst} across different \atbcr{3}{chips. Error bars show the range of \gls{hcfirst} across all tested rows in each chip.}}}
    \label{fig:chiphcfplot}
\end{figure}

\observation{It takes \emph{only} \param{14531} aggressor row activations to induce a RowHammer bitflip.}

The most RowHammer vulnerable DRAM row across all tested rows has an \gls{hcfirst} of \param{14531}. Causing this bitflip in a row in Chip 5 takes us \param{1.3} milliseconds. 

\observation{\gls{hcfirst} varies across chips and there are rows in every chip that exhibit relatively small \gls{hcfirst} values.}

There is variation in \gls{hcfirst} distributions between different chips for the same data pattern. For example, the mean \gls{hcfirst} value for Chip 5 is \param{10.59\%} higher than the mean \gls{hcfirst} value for Chip 2 using the Rowstripe0 data pattern. The other tested chips (Chips 0-4) display similar minimum \gls{hcfirst} values. It takes \param{18087}, \param{16611}, \param{15500}, \param{17164}, and \param{15500} aggressor row activations to induce the first RowHammer bitflip in Chips 0-4, respectively. 


\take{HBM2 chips exhibit different levels of RowHammer vulnerability in terms of mean \gls{ber} (up to \param{0.49} \revb{percentage points} \atbcr{1}{difference}) and minimum \gls{hcfirst} (up to \param{3556} \atbcr{1}{difference}).}
\label{take:rowhammer-change-chip}



\subsection{{RowHammer Across Channels}}

\figref{fig:berplot} shows the distribution of \gls{ber} (y-axis) across different DRAM rows for a given data pattern (x-axis) in a channel (color-coded).

\begin{figure}[th]
    \centering
    \includegraphics[width=\linewidth]{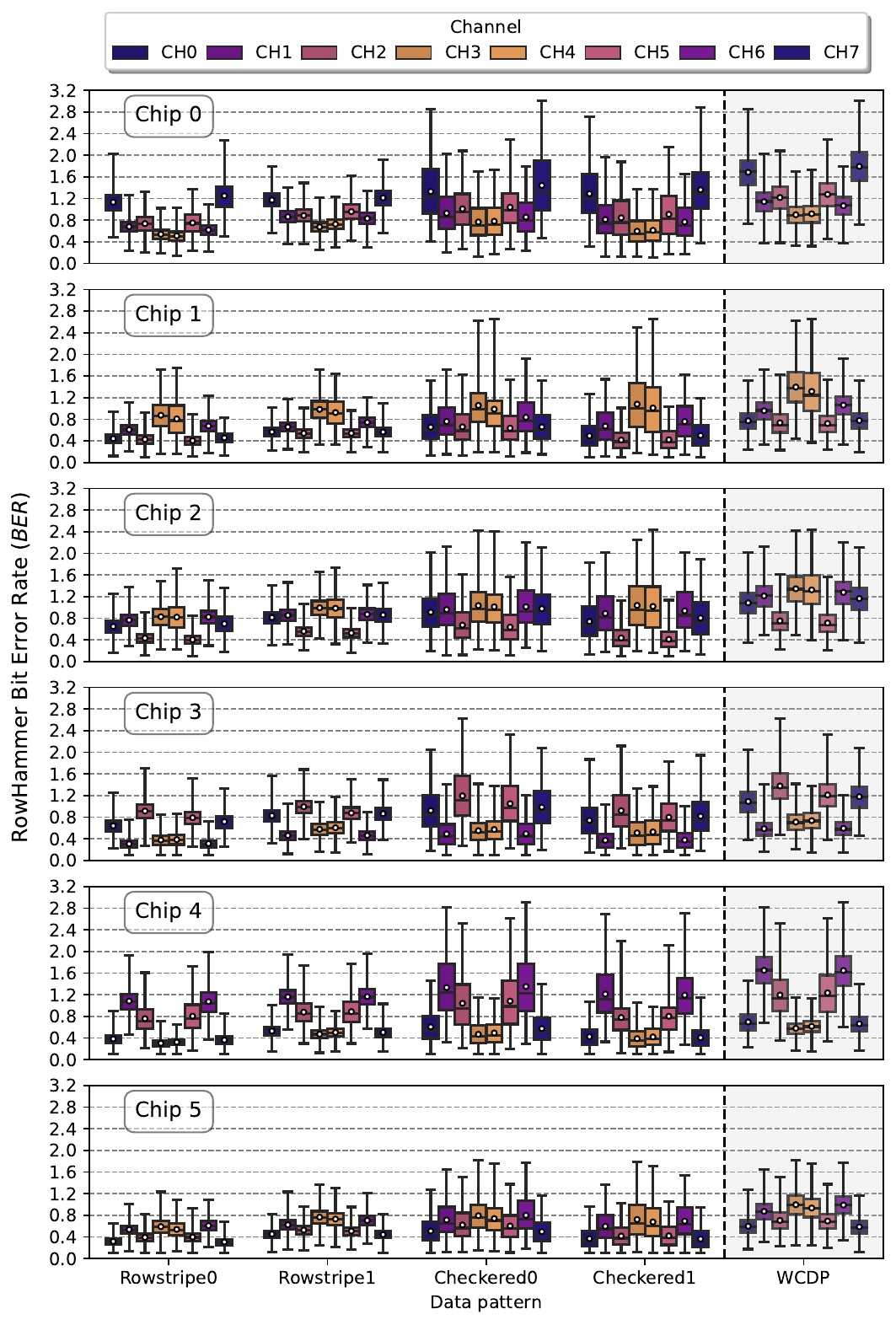}
    \caption{\Glsfirst{ber} across different DRAM \atbcr{3}{channels}. Error bars show the range of \gls{ber} across rows \atbcr{3}{in each channel}.}
    \label{fig:berplot}
\end{figure}

\observation{\revcommon{E}ach tested DRAM row across all tested HBM2 channels \revcommon{exhibits RowHammer bitflips}. The worst-case data pattern (WCDP) \gls{ber} across all tested rows in all chips can reach \param{3.02\%} (i.e., \param{247} bitflips out of 8192 bits in a row).}
\label{obs:ber-high}


\observation{\gls{ber} varies across channels in a chip.}

For example, in Chip 0, channels CH0 and CH7 exhibit significantly {higher \gls{ber}} than other channels. CH7 (where we observe the highest mean \gls{ber}) has $\param{1.99}\times$ the \gls{ber} of CH3 (where we observe the lowest mean \gls{ber}) for WCDP. We observe that channels can be classified into groups of two based on the number of bitflips they exhibit. We highlight {these groups using different shades of the same color in~{\figref{fig:berplot}}}. {For example, channels CH3 and CH4 exhibit a similar \gls{ber} distribution across rows in every tested HBM2 chip.} We hypothesize that groups of channels are spread across different HBM2 DRAM dies. The difference in \gls{ber} across the groups of channels could be due to process variation \dsnadd{as a 3D-stacked chip is typically constructed by stacking DRAM dies that pass functionality and performance testing~\cite{farmahini2018challenges} (the difference in \gls{ber} across groups of two channels} is similar to how different DDR3/4 chips exhibit different RowHammer characteristics~\cite{kim2014flipping,kim2020revisiting,orosa2021deeper, park2016statistical,park2016experiments,yaglikci2022understanding,yaglikci2024svard}). 



\observation{The distribution of \gls{ber} across channels changes from chip to chip.}

The most RowHammer vulnerable channel (i.e., a channel with \omcr{3}{the highest} RowHammer \gls{ber}) is \emph{not} necessarily the same across every chip.
{For example, channels CH0 and CH7 have the highest average \gls{ber} in Chip 0, whereas channels CH3 and CH4 have the highest average \gls{ber} in Chip 1 for WCDP. We hypothesize that this difference could be due to the effects of process variation across HBM2 chips and dies inside chips.} 

\observation{\atbcr{3}{The mean} \gls{ber} \atbcr{3}{across all rows in each channel has a wider distribution than the mean} \gls{ber} \atbcr{3}{across all rows in each chip}.}

For example, the difference between the maximum and the minimum mean \gls{ber} across all rows in each channel in Chip 4 is \param{0.88} \revb{pp}, whereas the difference between the maximum and the minimum mean \gls{ber} across all rows in each chip is \param{0.38} \revb{pp} (see \figref{fig:berplot}), for the Checkered0 data pattern. {This observation holds for all tested data patterns and chips except Chip 5. Chip 5 has a smaller difference between the maximum and the minimum mean \gls{ber} across all rows in each channel than the difference across all rows in each chip.} \revlabel{Rev.A-Q1}\reva{Manufacturing process variation inherent to 3D die stacking process~\cite{farmahini2018challenges} could explain this observation. 3D die stacking could alter the RowHammer vulnerability characteristics (e.g., \gls{ber}) of DRAM dies that make up a chip stack. We hypothesize that this alteration is made in a way that widens the \gls{ber} distribution across 3D-stacked dies in an HBM2 chip.}



\figref{fig:hcfplot} shows the distribution of \gls{hcfirst} (y-axis) across different DRAM rows for a given data pattern (x-axis) in a channel (color-coded).

\begin{figure}[th]
    \centering
    \includegraphics[width=\linewidth]{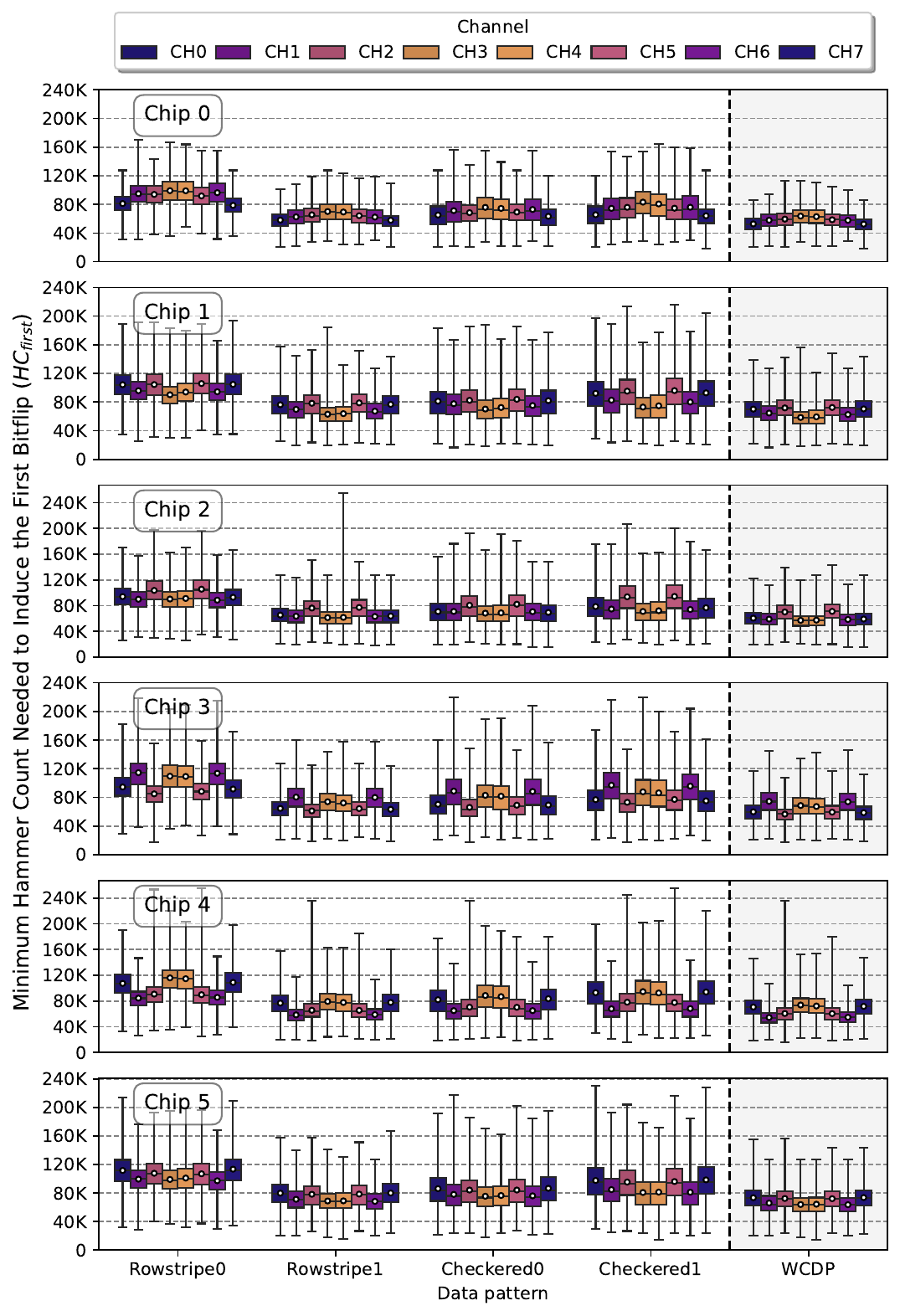}
    \caption{\Glsfirst{hcfirst} across different DRAM \atbcr{3}{channels.}}
    \label{fig:hcfplot}
\end{figure}


\observation{Different channels exhibit different \nrh{} distributions. These distributions are affected by the data pattern.}

For example, channels \param{CH3} and \param{CH4} in {chip 1} contain more rows with smaller \nrh{} values than other channels. Because these channels also exhibit more RowHammer bitflips than other channels {in chip 1} (see \figref{fig:berplot}), we hypothesize that these channels belong \atbcr{3}{to} the die with the worst RowHammer vulnerability across all dies. 
{The {median} \gls{hcfirst} for Rowstripe0 and Rowstripe1 in channel \atbcr{1}{CH0} {in chip 1} are \param{103905} and \param{75990}, respectively.} {Testing with different data patterns is necessary to assess the RowHammer vulnerability of {an} HBM2 DRAM chip, as no data pattern individually achieves the smallest \gls{hcfirst} or the {highest} \gls{ber} (\figref{fig:berplot}).}


\take{RowHammer \gls{ber} and \nrh{} vary between different 3D-stacked channels in a chip and with the data patterns in aggressor and victim rows. \atbcr{3}{The mean \gls{ber} across all rows in each channel has a wider distribution than the mean \gls{ber} across all rows in each chip.}}
\label{take:rowhammer-change-channel}

\figref{fig:beracrossrowsplot} shows the \gls{ber} \dsnadd{over} each DRAM row in a bank when we use the worst-case data pattern (WCDP) to initialize the rows. Each color-coded \gls{ber} curve represents the \gls{ber} for rows in three different channels. The {shaded regions} indicate variable-sized \emph{subarray} boundaries.\atbcrcomment{3}{RowClone results are consistent with our rev. engr. results.}\footnote{{We reverse engineer subarray boundaries by performing single-sided RowHammer~\cite{kim2014flipping,kim2020revisiting} that induces bitflips in \emph{only one} of the victim rows if the aggressor row is at the edge of a subarray. We find that a subarray contains either 832 (SA X in \figref{fig:beracrossrowsplot}) or \param{768} (SA Y in \figref{fig:beracrossrowsplot}) DRAM rows.}} For example, the green highlighted subarray (SA X) comprises 832, and the blue highlighted subarray (SA Y) comprises 768 DRAM rows. 

\begin{figure}[h]
    \centering
    \includegraphics[width=\linewidth]{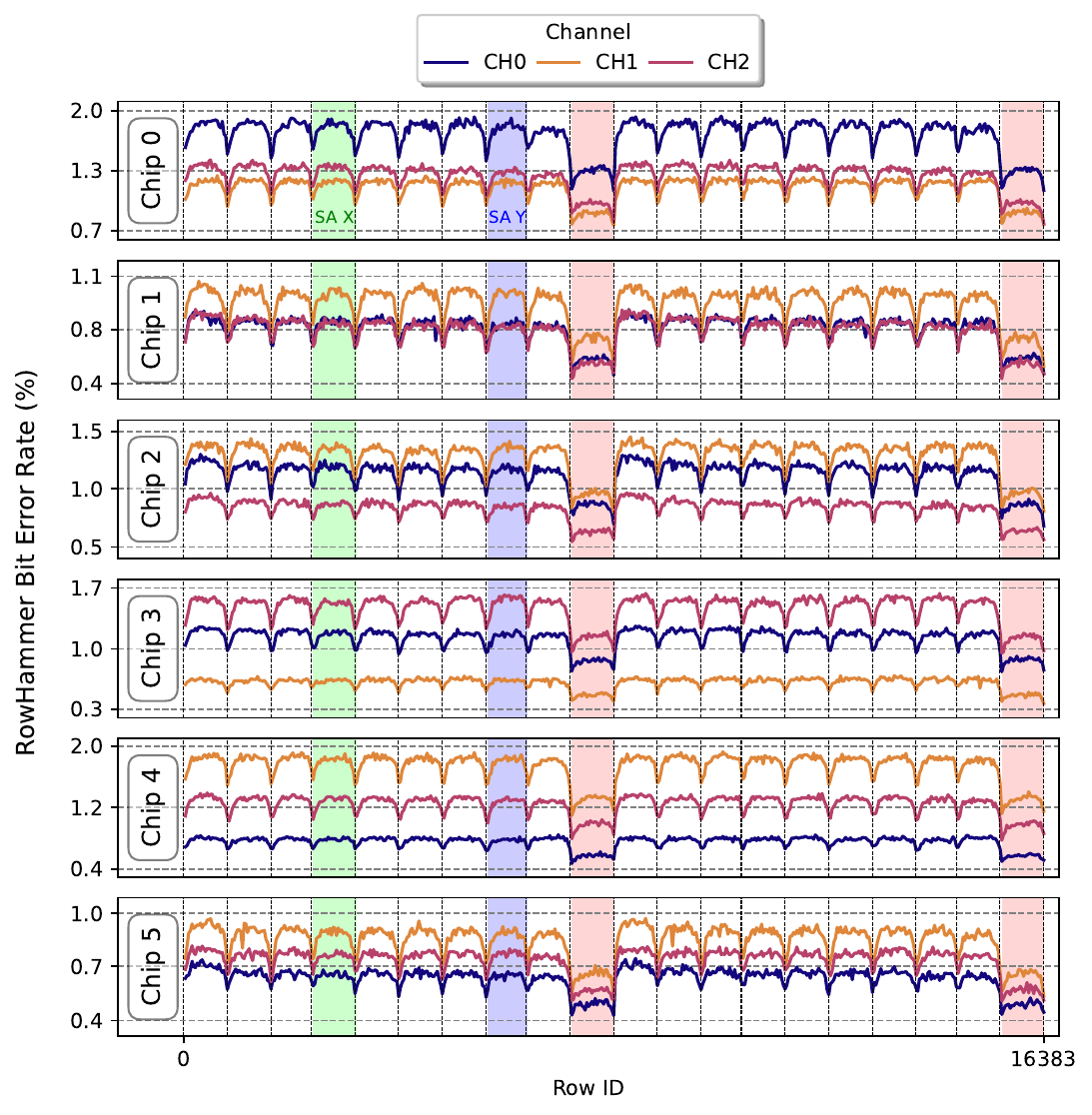}
    \caption{\Glsfirst{ber} for different rows across a bank in different channels. \atbcr{4}{Dashed vertical lines indicate subarray boundaries.}}
    \label{fig:beracrossrowsplot}
\end{figure}


\observation{\gls{ber} periodically increases and decreases across DRAM rows.}

{This observation is consistent in each tested chip.} \gls{ber} is higher in the middle of a \emph{subarray} and lower towards either end. {We hypothesize that the increasing and decreasing pattern results from the structure of the local DRAM array. \atbcrcomment{3}{Mfr. H. in Svard has a similar looking BER behavior. We do not have a good hypothesis as to why BER increases towards the middle.}For example, the RowHammer vulnerability of a row could increase with the row's distance from the row buffer.} 

\observation{The last {and the middle} subarrays {in a} bank exhibit relatively low \gls{ber}.}

The last {and the middle} subarrays {in a} bank ({highlighted with red color in~{\figref{fig:beracrossrowsplot}}}), which contain the last {and the middle} 832 DRAM rows, exhibit significantly lower BER than the other subarrays. We hypothesize that the{se subarrays} exhibit smaller \gls{ber} due to the micro-architectural characteristics of the DRAM bank. \revlabel{Rev.C-C2}{\revc{First}, assuming that proximity to the shared I/O circuitry on the DRAM die affects the RowHammer vulnerability of a subarray, the last {and the middle} subarray{s} might be placed near this shared I/O circuitry~\cite{jun2017hbm}. \revc{Second, an edge subarray (a subarray at either edge of \omcr{3}{a} bank) typically harbors different design characteristics than other subarrays~\cite{nam2023xray,kim2021imbalance}. \omcr{3}{Edge} subarrays (e.g., the last and the middle subarrays) in the tested chips could exhibit a design characteristic that results in their DRAM wordlines to be driven to a smaller wordline voltage ($V_{PP}$) than other subarrays' wordlines', which could explain the relatively low \gls{ber} values in the middle and last subarrays~\cite{yaglikci2022understanding}.} \atbcr{1}{Third, the tested data pattern may not be replicated in the same way in edge subarrays due to inaccessible ``dummy bitlines''~\cite{nam2023xray}\atbcrcomment{3}{The columns exist as a side-effect of open-bitline architecture. The X-ray paper~\cite{nam2023xray} describes this architecture in Fig 3. The edge subarrays somehow cooperate to serve the same row size even though the two edge subarrays seemingly have the same number of rows. This is a weak hypothesis we developed with Michele and Haocong. I want to keep it weak, ideally.} and thus the data pattern in the last and middle subarrays could be inducing a smaller read disturbance effect.}}

\take{A subset of HBM2 rows (the middle and the last 832 rows) are significantly more RowHammer resilient than the other rows in an HBM2 channel.}
\label{take:subarray}

\subsection{{RowHammer Across Banks and Pseudo Channels}}

To investigate the variation in the RowHammer vulnerability across HBM2 banks and pseudo channels, we measure \gls{ber} on 300 rows from 256 banks {in chip 0}.\footnote{First, middle, and last 100 rows in each of the 256 banks spread across eight channels and two pseudo channels.}
\figref{fig:bankvar} compares different banks' \gls{ber} distributions across channels (color) and {pseudo channels} (marker style) in terms of the coefficient of variation (CV)\footnote{Coefficient of variation is the standard deviation of a distribution normalized to the mean.} (x-axis) and mean BER (y-axis). We draw one marker for each bank. At a high level, a marker close to the y-axis (e.g., the leftmost markers) indicates that the variation in \gls{ber} across rows in that bank is smaller. A marker close to the x-axis (e.g., the bottommost markers) suggests that the mean \gls{ber} across rows in that bank is smaller.

\begin{figure}[!h]
    \centering
    \includegraphics[width=\linewidth]{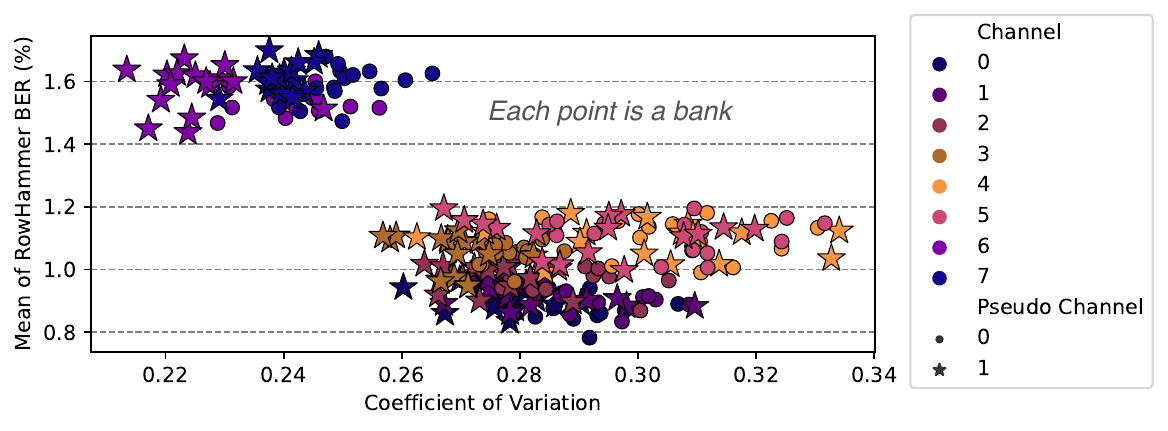}
    \caption{\gls{ber} variation across banks. Each bank is represented by the average BER (y-axis) and the coefficient of variation in BER (x-axis) across \omcr{3}{all} rows within the bank.}
    \label{fig:bankvar}
\end{figure}


\observation{RowHammer \gls{ber} varies across banks and pseudo channels. The \gls{ber} variation across banks is dominated by variation across channels.}

For example, there is up to {\param{0.23} \revb{pp}} difference in mean \gls{ber} across banks in channel \param{7}. 
The markers follow a bimodal distribution (i.e., the markers are clustered around two points in the plot). The two clusters indicate that 1)~a bank with a higher mean \gls{ber} across its rows also has a smaller deviation (i.e., smaller coefficient of variation) from \atbcr{3}{the average} \gls{ber} across its rows and 2)~a bank with a smaller mean \gls{ber} across its rows also has a larger deviation from \atbcr{3}{the average} \gls{ber} across its rows. \omcrcomment{3}{hard to understand. also the terms and writing here does not seem correct.} 
Banks in different channels tend to have a larger \gls{ber} difference than banks in the same channel (\figref{fig:berplot}), indicating that testing different channels is more important than testing different banks or pseudo channels in providing a comprehensive understanding of the RowHammer vulnerability in HBM2 DRAM chips.

\take{RowHammer \gls{ber} varies \omcr{3}{across} pseudo channels and banks. This variation is less prominent than the \gls{ber} variation between channels.}
\label{take:rowhammer-change-bank-pseudo-channel}
\noindent

\subsection{\revc{The Effect of Aging}}
\revc{A DRAM row’s RowHammer vulnerability can change\revlabel{Rev.C-Q1} over time. A rigorous characterization study on many HBM2 
chips over a large timespan is required to comprehensively demonstrate the effects of aging. Due to time and space 
limitations, we leave such \omcr{3}{a study} for future work while presenting a preliminary analysis using four HBM2 chips (\omcr{3}{Chips} 2-5, which have the same estimated age) as our best effort. We repeat our \gls{ber} experiments for 3072 rows in 3 channels after 7 months of 
keeping the HBM2 chips powered on.
\figref{fig:ber-aging} (left) shows the distribution of a row's \gls{ber} after aging (New \gls{ber}) over its \gls{ber} before aging (Old \gls{ber}) for the 
Checkered1 data pattern. \figref{fig:ber-aging} (right) shows the distribution of a row's \gls{ber} before aging over its \gls{ber} 
after aging. The left subplot shows only the distribution for the rows whose \atbcr{4}{\gls{ber}s} increase after aging and the right subplot 
shows the distribution for the rest of the rows. The x-axis is marked with percentiles ranging from P1 to P99.}

\begin{figure}[h]
    \centering
    \includegraphics[width=\linewidth]{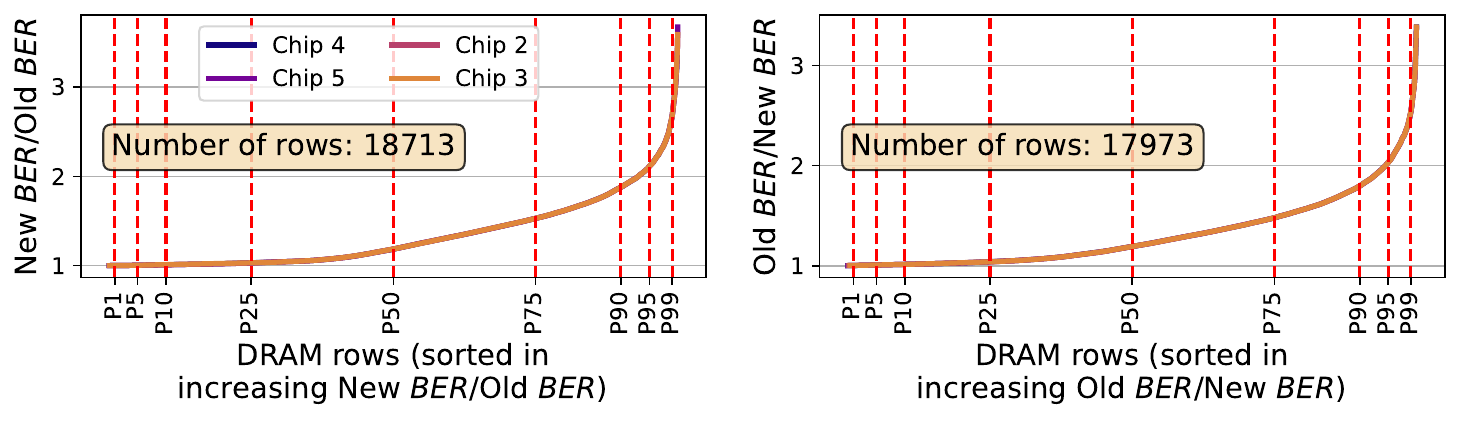}
    \caption{\revc{Distribution of the deviation of \gls{ber} after aging from \gls{ber} before aging across tested DRAM rows.}}
    \label{fig:ber-aging}
\end{figure}

\observation{\revc{The \gls{ber} of a DRAM row changes after aging. A larger fraction of the tested DRAM rows have higher \gls{ber} after aging.}}

\revc{18713 and 17973 of the tested DRAM rows have a higher New \gls{ber} and a lower New \gls{ber}, respectively (we omit 178 outlier rows from the figure). We observe similar distributions for all tested HBM2 chips.}

\section{RowHammer's Sensitivity to Hammer Count}
\label{sec:hcnth}

We analyze the number of aggressor row activations \omcr{3}{(hammer count)} needed to induce up to 10 bitflips in a DRAM row. Our \nrh{} analysis \omcr{3}{(\secref{sec:spatial-variation-analysis})} already \omcr{3}{examined} the hammer count to induce one (\emph{the first}) bitflip in a row. We use the same naming convention used with \nrh{} to refer to these 9 new hammer counts that we determine in this analysis. For example, we call the hammer count to induce \emph{the second} bitflip $HC_{second}$ and \emph{the tenth} bitflip $HC_{tenth}$. We report \omcr{3}{each of} the 9 new hammer counts \omcr{3}{as a value normalized to} \nrh{}. For example, if a row's \nrh{} is \param{10} and its $HC_{second}$ normalized to \nrh{} is \param{2}, the absolute $HC_{second}$ of the row is \param{20}. We record the hammer counts to induce up to 10 bitflips in 32 rows from each of the beginning, middle, and end of one bank in two channels (that exhibit the smallest \nrh{} across all channels) in every HBM2 chip. 
\figref{fig:hcnth-plot} plots the distribution of all 10 hammer counts (x-axis) normalized to \nrh{} (y-axis) for all tested DRAM rows (1152 such rows across all chips).


\begin{figure}[h]
    \centering
    \includegraphics[width=\linewidth]{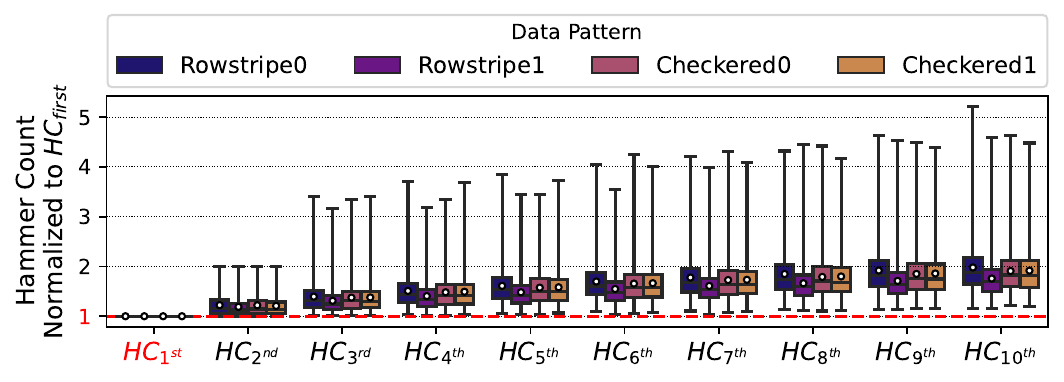}
    \caption{Distribution of hammer counts (y-axis) to induce up to 10 bitflips in a DRAM row (x-axis), normalized to the row's \gls{hcfirst}.}
    \label{fig:hcnth-plot}
\end{figure}


\observation{\omcr{3}{H}ammer count \omcr{3}{needed} to induce up to 10 bitflips in a row significantly varies between rows.}

The hammer count to induce up to 10 bitflips in a row can be as small as \param{1.15}$\times{}$ and as large as \param{5.22}$\times{}$ the \nrh{} of the DRAM row. Fewer than 2$\times{}$ \nrh{} hammers are enough to induce 10 bitflips in a DRAM row on average across all tested DRAM rows. For example, an average DRAM row's $HC_{second}$, $HC_{fourth}$, $HC_{eighth}$, and $HC_{tenth}$ are \param{1.19}$\times{}$, \param{1.41}$\times{}$, \param{1.66}$\times{}$, and \param{1.76}$\times{}$ that of the row's \nrh{} for the Rowstripe1 data pattern. 

\observation{The hammer counts to induce up to 10 bitflips are moderately affected by data patterns.}

For example, the difference between the largest (Rowstripe0) and the smallest (Rowstripe1) mean normalized $HC_{tenth}$ is \param{12.59\%}. The variation in normalized hammer count across data patterns resembles the variation in \nrh{} across data patterns (see \figref{fig:chiphcfplot}).




\figref{fig:hcfirst-vs-hctenth} plots the additional hammer count \omcr{3}{(over \nrh{})} to induce the $10^{th}$ bitflip (y-axis) for all tested DRAM rows whose \nrh{} values are depicted on the x-axis. We compute the additional hammer count for a row as the row's $HC_{tenth}-HC_{first}$. \figref{fig:hcfirst-vs-hctenth} shows one subplot for each of the 6 tested HBM2 chips.

\begin{figure}[h]
    \centering
    \includegraphics[width=\linewidth]{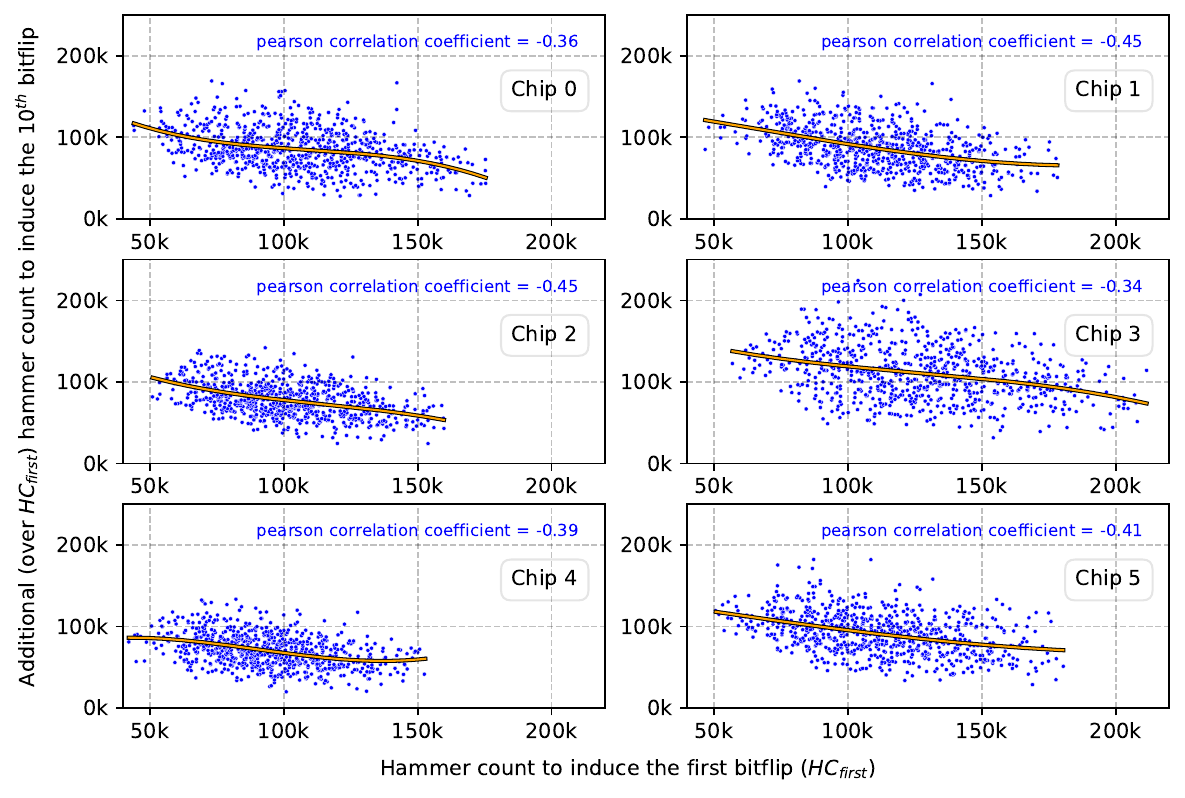}
    \caption{Additional (over \nrh{}, x-axis) hammer count needed to induce the \atbcr{1}{tenth} bitflip (y-axis) for each tested DRAM row in each tested chip (labeled Chip 0 to Chip 5 in the figure). We plot a polynomial curve fit (orange curve) for each distribution to highlight the decreasing additional hammer count trend with increasing \nrh{}.}
    \label{fig:hcfirst-vs-hctenth}
\end{figure}


\observation{It takes fewer additional hammer counts (over \nrh{}) to induce the $10^{th}$ RowHammer bitflip for a DRAM row with a large \nrh{} compared to a DRAM row with a small \nrh{}.}

We observe that increasing \nrh{} is correlated with decreasing \emph{additional} hammer count to induce the $10^{th}$ bitflip. We compute the {P}earson correlation coefficient for each distribution to quantify the correlation. We conclude that increasing \nrh{} is \emph{moderately} correlated with decreasing additional hammer count to induce the $10^{th}$ bitflip, based on the {weakest} (\param{$-0.34$}) and the {strongest} (\param{$-0.45$}) {P}earson correlation we observe across distributions for each chip (displayed on each subplot).

\take{It can take fewer aggressor row activations to induce multiple (e.g., 10) bitflips in a DRAM row if it takes many activations to induce the first bitflip in the row.}
\label{take:hcnth}
\vspace{3mm}

\revb{\figref{fig:hcfirst-row-distribution} shows the distribution\revlabel{Rev.B-Q1} of the \omcr{3}{maximum change} of \gls{hcfirst} from the minimum observed \gls{hcfirst} for a DRAM row 
over 50 experiment iterations using the Rowstripe0 data pattern. The x-axis shows all tested rows (768 rows from channel 0 in every tested HBM2 chip), sorted by increasing \omcr{3}{maximum change} of \gls{hcfirst} and marked with percentiles ranging from P1 to P99.\footnote{\revb{To maintain a reasonable testing time, we investigate the \omcr{3}{maximum change} of \gls{hcfirst} using one data pattern and 4608 DRAM rows. We leave the detailed analysis of the \omcr{3}{maximum change} of \gls{hcfirst} that covers more data patterns, HBM2 components (e.g., channels, pseudo channels, banks, rows), and access patterns for future work.}}} 

\begin{figure}[h] 
    \centering
    \includegraphics[width=0.8\linewidth]{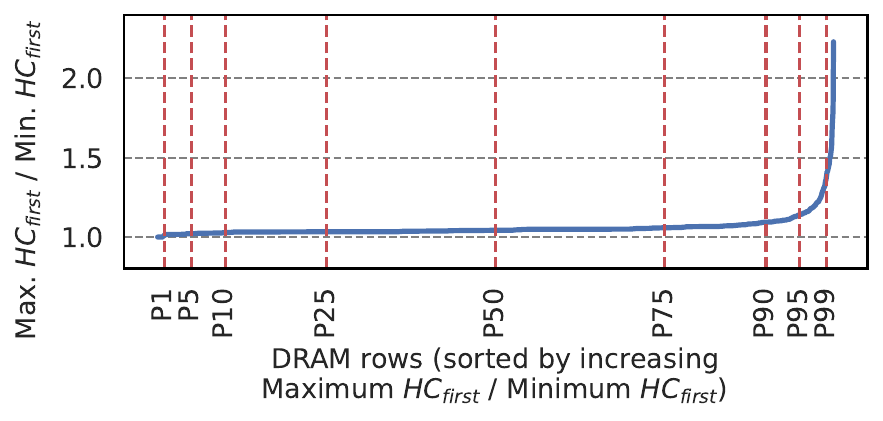}
    \caption{\revb{Distribution of the maximum observed \gls{hcfirst} over the minimum observed \gls{hcfirst} for a DRAM row \omcr{3}{across experiments}, across all tested 4608 DRAM rows for the Rowstripe0 data pattern.}}
    \label{fig:hcfirst-row-distribution}
\end{figure}

\revb{We make two key observations. First, the \gls{hcfirst} of a DRAM row can significantly \omcr{3}{change} from the minimum observed \gls{hcfirst} across experiment iterations (i.e., over time). The greatest \omcr{3}{change} that we observe for a row is 2.23$\times{}$ (the row's maximum observed \gls{hcfirst} is 2.23$\times{}$ of its minimum observed \gls{hcfirst}). Second, the majority of rows experience a small \gls{hcfirst} \omcr{3}{change}. For example, 90\% of all tested rows experience an \gls{hcfirst} \omcr{3}{change} smaller than 1.09$\times{}$.}


\section{{RowHammer and RowPress's Sensitivities to\\Aggressor Row On Time}}
\label{sec:taggon}

With aggressive technology node scaling, DRAM suffers from worsening read disturbance effects. One prominent example \omcr{3}{of} read disturbance in DRAM is RowHammer, which we \omcr{3}{have already} extensively characterize\omcr{3}{d} in our HBM2 chips. We also investigate the characteristics of another widespread read disturbance effect called RowPress, recently experimentally demonstrated in real DDR4 chips~\cite{luo2023rowpress}. RowPress \omcr{3}{is the phenomenon} that keeping an aggressor row open for a long period of time (i.e., a large aggressor row on time, $t_{AggON}$) induces bitflips in physically nearby DRAM rows with orders of magnitude smaller hammer counts compared to a traditional RowHammer access pattern which keeps the aggressor row open for a short period of time. We provide the first extensive analysis of RowPress in six real HBM2 chips.


\figref{fig:ber-taggon} depicts how \gls{ber} (y-axis) varies with
increasing $t_{AggON}$ (x-axis) across \param{the first, middle, and last 128 rows in one DRAM bank} for 8 channels when we use a hammer count of 150K (i.e., activate each aggressor row 150K times) and the Checkered0 data pattern. We plot \atbcr{1}{the results for} four relatively small (left subplots) and two relatively large (right subplots) $t_{AggON}$ values. The minimum $t_{AggON}$ is the $t_{RAS}$ timing parameter~\cite{jedec2021hbm,oconnor2021thesis}. We choose two relatively large $t_{AggON}$ values of interest as $t_{REFI}$, the average interval between two successive periodic refresh commands, and as $9*t_{REFI}$, the maximum interval between two subsequent periodic refresh commands (i.e., the maximum time a row can remain open according to the HBM2 standard)~\cite{jedec2021hbm}. 

\noindent
\ext{\textbf{Factoring in Retention Errors.} $t_{AggON}$ values above \SI{116.0}{\nano\second} combined with the $150K$ hammer count results in experiment times that are longer than the \SI{32}{\milli\second} refresh window \atbcr{3}{(e.g., $150K$ hammers at $t_{AggON}=35.1\mu{}s$ takes 10.53 seconds)}, where DRAM cells can exhibit \emph{retention failures}. Because we want to analyze \emph{only} the read disturbance bitflips, we remove bitflips that are caused by retention failures \atbcr{3}{(\param{0.109\%} of all bitflips we observe)} from the set of bitflips we observe at every such $t_{AggON}$. To do so, we perform retention profiling experiments where we initialize a tested DRAM row with the Checkered 0 data pattern, wait for the tested retention time to pass (\param{\SI{34.8}{\milli\second}}, \param{\SI{1.17}{\second}}, and \param{\SI{10.53}{\second}} for \gls{taggon}=\SI{116.0}{\nano\second}, \SI{3.9}{\micro\second}, and \SI{35.1}{\micro\second}, respectively), and read back the DRAM row. We perform each such experiment 5 times and consider a cell to exhibit retention failures at the tested retention time if the cell exhibits a retention failure in any of the 5 experiments. The average retention \gls{ber} for \param{\SI{34.8}{\milli\second}}, \param{\SI{1.17}{\second}}, and \param{\SI{10.53}{\second} is 0\%, 0.013\%, and 0.134\%, respecitvely.}}

\noindent
\revlabel{Rev.B-Q2}\revb{\textbf{$t_{AggON}$ in commodity systems.} In general, the time a DRAM row is kept open depends on two factors: 1) $t_{REFI}$ in the DRAM specification, and 2) the memory request scheduling algorithm implemented in the memory controller. For HBM2, the default $t_{REFI}$ is $3.9\mu{}s$. The memory controller in a commodity CPU or a GPU may delay up to 8 periodic refresh commands and thereby keep a row open for as long as 9*$t_{REFI}$ ($35.1\mu{}s$). The memory request scheduling algorithm is typically \emph{not} disclosed by system manufacturers. However, it is reasonable to assume that a high-performance scheduling algorithm implemented in a commodity CPU or GPU's memory controller would keep a DRAM row open at least for as long as required to serve a regular memory access pattern that ``streams’’ through a row (i.e., accesses every cache line in the row one by one)~\cite{jog2015anatomy,nordquist2009apparatus,jog2016exploiting,ausavarungnirun2012staged}, to exploit row buffer locality and thus maximize memory throughput. We estimate the time it takes to stream through a row as \atbcrcomment{3}{The row is much smaller (8X) than DDR4.}$128.0ns$ (32*$tCCD\_L$) based on the HBM2 timing parameter values depicted in~\cite{oconnor2021thesis}.}

\begin{figure}[!ht]
    \centering
    \includegraphics[width=\linewidth]{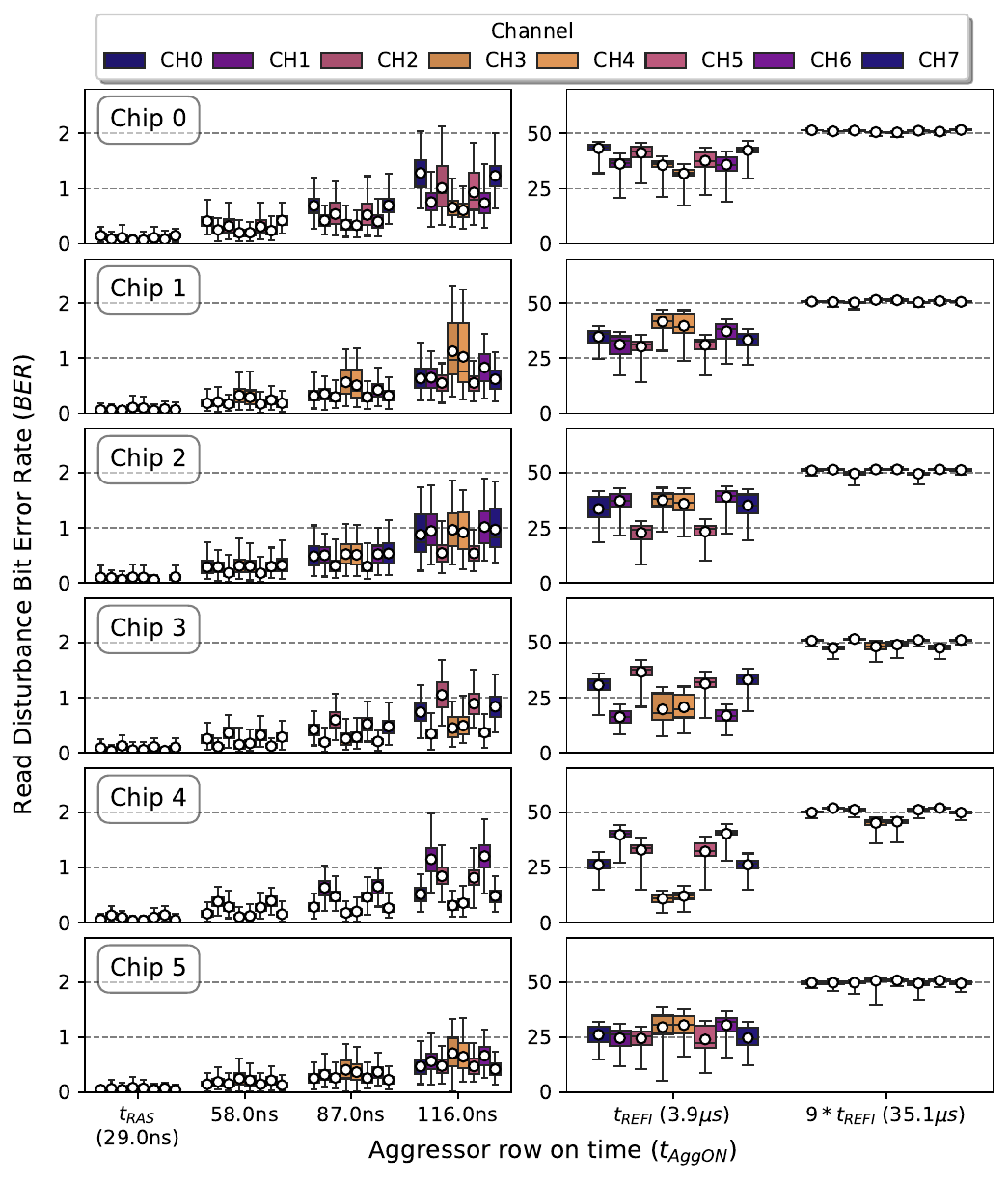}
    \caption{{\gls{ber} with increasing \gls{taggon}.}}
    \label{fig:ber-taggon}
\end{figure}

\observation{RowPress \gls{ber} in each chip consistently increases with \gls{taggon} in all tested channels.}

The average \gls{ber} across every channel in every chip is \param{0.08}\%, \param{0.24}\%, \param{0.40}\%, \param{0.73}\%, \param{31.00}\%, and \param{50.35}\% at \gls{taggon} of \param{\SI{29.0}{\nano\second}}, \param{\SI{58.0}{\nano\second}}, \param{\SI{87.0}{\nano\second}}, \param{\SI{116.0}{\nano\second}}, \param{\SI{3.9}{\micro\second}}, and \param{\SI{35.1}{\micro\second}}, respectively. 

\observation{Channels with high \gls{ber} at low \gls{taggon} values tend to have high \gls{ber} also at high \gls{taggon} values.}

For example, \atbcr{3}{CH} 1 in Chip 3 has the highest mean \gls{ber} across all tested \gls{taggon} values. We observe that all \gls{ber} values converge to around $50\%$ for the \gls{taggon} of \SI{35.1}{\micro\second} across all tested channels. We hypothesize that this is due to the combined effect of 1)~the data pattern that we use which initializes a victim DRAM row with alternating \emph{logic-1} and \emph{logic-0} (i.e., 10101010...) and 2)~RowPress causing bitflips from \emph{logic-1} to \emph{logic-0} more frequently than from \emph{logic-0} to \emph{logic-1} (as demonstrated in~\cite{luo2023rowpress} for a wide variety of DDR4 chips).

{\figref{fig:hcfirst-taggon} depicts how the \gls{hcfirst} values (y-axis) across \param{384 tested DRAM rows in a channel} change with increasing \gls{taggon} (x-axis) for \param{3} channels when we use the Checkered0 data pattern. From left to right, 1) the default \gls{tras} value (\SI{29.0}{\nano\second}), 2) \gls{trefi}, the average time interval between two successive periodic refresh commands, 3) $9*t_{REFI}$, the maximum time a row can remain open according to the HBM2 specification, and 4) half the \gls{trefw}~\cite{jedec2021hbm} (such that each aggressor row can be activated once in a \gls{trefw}). {\figref{fig:hcfirst-taggon} shows} one subplot for every tested HBM2 chip. The grey boxes indicate the number of rows we show in each subplot. We only show rows for which we observe the first \atbcr{3}{read-disturb} bitflip in a refresh window (under \SI{32}{\milli\second}) at every tested \gls{taggon} value. We display the results for four \gls{taggon} values on the x-axis.}

\begin{figure}[!ht]
    \centering
    \includegraphics[width=\linewidth]{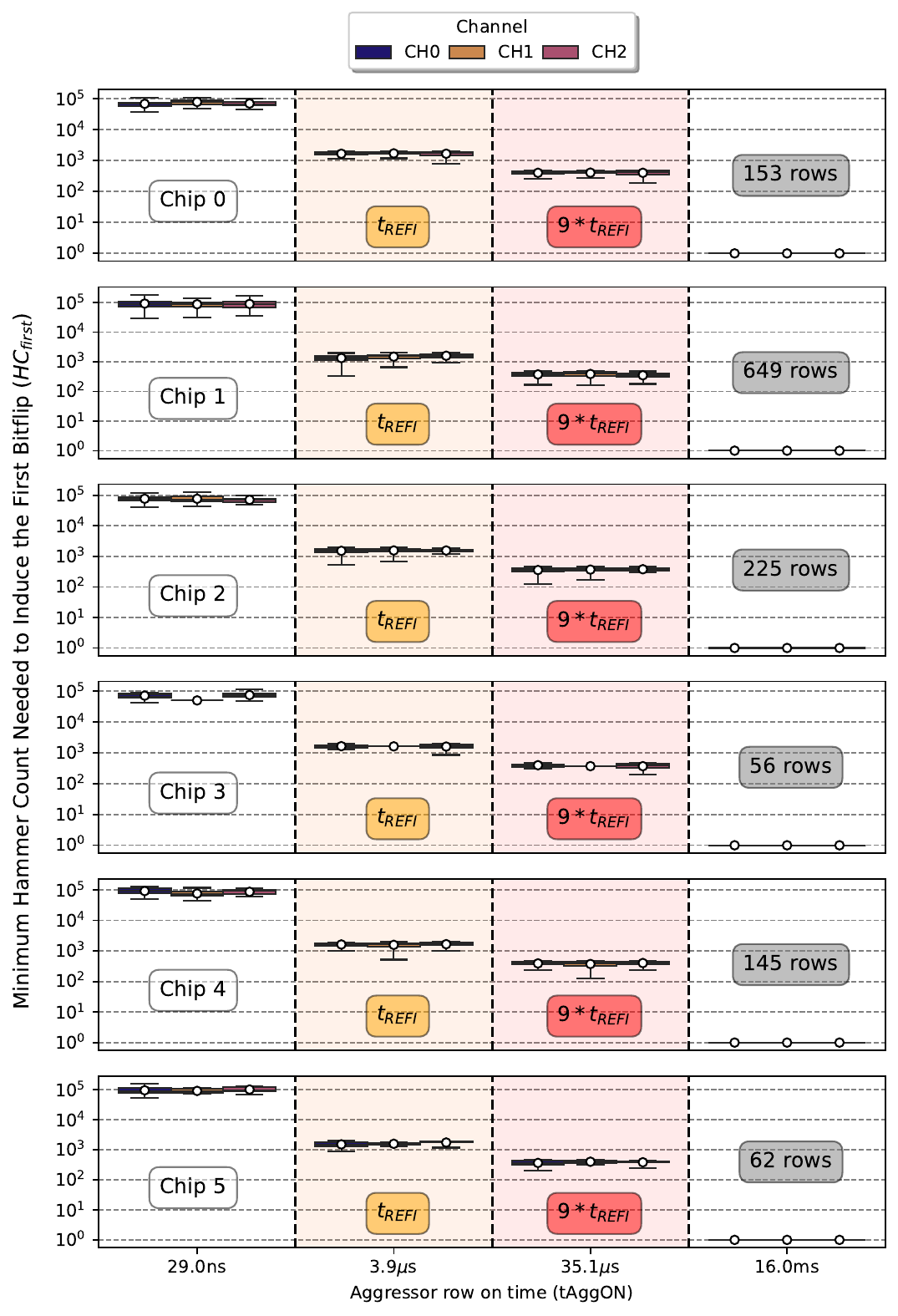}
    \caption{{\gls{hcfirst} with increasing \gls{taggon}.}}
    \label{fig:hcfirst-taggon}
\end{figure}

\observation{As the aggressor row remains open longer (i.e., as \gls{taggon} increases), DRAM rows experience bitflips \omcr{3}{at smaller} hammer counts.}

This observation is consistent across the three tested HBM2 channels. {The average (minimum) \gls{hcfirst} values across all chips are \param{83689} (\param{29183}), \param{1519} (\param{335}), \param{376} (\param{123}), and \param{1} (\param{1}), for the four tested \gls{taggon} values, respectively.}

\take{The read disturbance vulnerability of tested HBM2 chips worsens (i.e., \gls{ber} increases and \gls{hcfirst} reduces) with increasing \gls{taggon}.}
\label{take:taggon}
\vspace{1mm}

{Our observations are in line with prior \atbcr{4}{works} that investigate the effects of \gls{taggon} in real DDR4 chips~\cite{luo2023rowpress, orosa2021deeper}.}

\section{In-DRAM RowHammer Defenses}
\label{sec:uncovering}

To prevent RowHammer bitflips, DRAM manufacturers equip their chips with a mitigation mechanism broadly referred to as Target Row Refresh (TRR)~\cite{hassan2021utrr,frigo2020trrespass,micron2018ddr4trr}. Unfortunately, manufacturers do \emph{not} disclose the operational principles \omcr{3}{or implementations} of proprietary TRR versions (e.g., in DDR4). At a high level, TRR \omcr{3}{1)} identifies potential aggressor rows as the memory controller issues activate commands to the DRAM chip and \omcr{3}{2)} preventively refreshes their victim rows {when the memory controller issues a periodic $REF$ command}~\cite{hassan2021utrr,frigo2020trrespass}. 


We demonstrate that a tested HBM2 chip ({Chip 0}) implements a form of proprietary TRR (similar to the ones used in DDR4).\textsuperscript{\atbcr{3}{\ref{foot:trr}}} {We analyze the TRR mechanism and craft a specialized \revcommon{access} pattern that bypasses the TRR mechanism and {induces} RowHammer bitflips.\atbcr{3}{\footnote{\revlabel{Rev.CC-2}\revcommon{We note that the profiling required to perform system-level RowHammer attacks~\exploitingRowHammerAllCitations{} is \emph{not} as extensive as what we do in this work and can be done \atbcr{3}{using user-level programs}, as shown in various works (e.g.,~\cite{deridder2021smash,frigo2020trrespass,jattke2022blacksmith}).}}}\atbcrcomment{3}{The out of place sentence is to address a DSN reviewer comment. I put it as a footnote now.}}

\noindent
\textbf{Methodology.}
{We use U-TRR's~\cite{hassan2021utrr} methodology to uncover the proprietary TRR mechanism. {The key idea of this methodology is to use retention failures as a side channel to infer whether or \emph{not} TRR refreshes a DRAM row. 
{Our analysis consists of two steps. {First, we identify multiple DRAM rows with similar retention times (e.g., two rows that can correctly retain data when they are \emph{not} refreshed for the same amount of time) by profiling DRAM rows for their retention times. We test all of the DRAM rows {in bank 0} for retention failures starting with a retention time of \SI{64}{\milli\second} with increments of \SI{64}{\milli\second}. We deem a row to have a retention time of {T} if any of the DRAM cells in the row exhibit a bitflip at a retention time of {T}. \atbcr{3}{We find the smallest T that causes at least one retention bitflip for each tested row.}}}
{Second, to understand if the TRR mechanism samples an aggressor row, we execute a \param{four}-step process: 1) We initialize DRAM rows that have a retention time of T (we call these rows \emph{side-channel rows}) and wait for T/2 \emph{without} refreshing these rows. 2) We activate each of the DRAM rows adjacent to the side-channel rows \emph{once}. We hypothesize that the TRR mechanism samples an activation to an adjacent row as an aggressor row activation. 3) We issue a $REF$ command to trigger the TRR mechanism. If the TRR mechanism \atbcr{3}{samples} any of the activated adjacent rows in step 2, we expect the TRR mechanism to refresh the side-channel rows. 4) We wait for T/2, read the data in the side-channel rows and check for any retention bitflips. The side-channel rows exhibit retention bitflips \emph{only} if they are \emph{not} refreshed by the TRR mechanism. We use this information to understand how the TRR mechanism works.}

{We {repeat} our experiment with various carefully crafted {RowHammer access patterns} to understand how the TRR mechanism tracks the {aggressor rows}. 

\observation{Every $17^{th}$ $REF$ command can perform a {TRR victim row refresh} (i.e., every $17^{th}$ $REF$ command is TRR-capable).}

\observation{The TRR mechanism refreshes both of the adjacent rows of an aggressor row.}

{If} TRR \atbcr{1}{identifies} row R as an aggressor row, it refreshes rows R+1 and R-1. 
{\omcr{3}{Obs. 20 \& 21}} resemble the TRR mechanism {employed in real DDR4 chips from} Vendor C in U-TRR~\cite{hassan2021utrr}. 

\observation{The first row that gets activated after a TRR-capable $REF$ is \emph{always} \atbcr{3}{identified} as an aggressor row by the TRR mechanism.}

\observation{The TRR mechanism records the activation count of activated rows and uses this record to \atbcr{1}{identify} if a row is an aggressor row.}

\atbcrcomment{3}{Obs 22 23 do not resemble Vendor C. We believe they are unique (except obs 23 is phrased generally, the detailed explanation of it does not resemble Vendor C).}

Between two $REF$ commands, if a row is activated more than half the total activation count, TRR \atbcr{1}{identifies} that row as an aggressor row. For example, if {we issue} 10 activations between two $REF$ commands, the {row that receives the first $ACT$ command} and the row {that receives} 5 {$ACT$ commands} are \atbcr{1}{identified} by the TRR mechanism.} \revlabel{Rev.E-Q1}\reve{We compare our findings on the HBM2 chip’s TRR mechanism to U-TRR’s findings~\cite{hassan2021utrr} in DDR4 chips. U-TRR does \emph{not} \omcr{3}{make} similar observations to Observations 22 and 23. Thus, the HBM2 chip likely implements a previously-unknown type of TRR.}

\take{An HBM2 chip implements a proprietary TRR mechanism that tracks aggressor rows and proactively refreshes their victim rows.}
\label{take:there-is-trr}
\vspace{1mm}

\noindent
\textbf{Bypassing the proprietary TRR mechanism.}
{Based on our {observations}, we craft a specialized \revcommon{access} pattern that bypasses the TRR mechanism and causes RowHammer bitflips. {We calculate} the total activation \atbcr{3}{count} budget (i.e., the maximum number of $ACT$ commands that the memory controller can issue) between two $REF$ commands as $\lfloor{}(tREFI-tRFC)/tRC\rfloor{} = 78$~\cite{jedec2021hbm,oconnor2021thesis} for the tested HBM2 chip and fully utilize the activation \atbcr{3}{count} budget in our \revcommon{access} pattern. The key idea of this \revcommon{access} pattern is to trick TRR into \emph{not} \atbcr{1}{identifying} a \emph{real} aggressor row by repeatedly accessing multiple \emph{dummy} rows many times.
{T}he \revcommon{access} pattern 1) activates dummy rows (we vary the number of dummy rows and the activation count of the dummy rows) and {2)} performs double-sided RowHammer {using two real aggressor rows}. {We create this \revcommon{access} pattern such that} the number of {real} aggressor activation{s does} \emph{not} exceed half of the total activation \atbcr{3}{count} budget.} We repeat the \revcommon{access} pattern $8205 * 2$ times (i.e., approximately for two $tREFW\!$, \SI{64}{\milli\second})
for each DRAM row in a bank so that we activate aggressor rows as many times as possible before each DRAM cell is refreshed.

We test all rows in a bank of the HBM2 chip using this \revcommon{access} pattern while obeying manufacturer-recommended timing parameters (e.g., {we issue a $REF$ command every} $tREFI$, $3.9\mu{}s$). \figref{fig:utrr} shows the distribution of \atbcr{3}{\gls{ber}} for different numbers of dummy rows (x-axis) and aggressor activation counts (different boxes). Since we have a total activation \atbcr{3}{count} budget of 78 and we utilize the whole budget for aggressor row and dummy row activations, the number of dummy row activations varies between boxes displayed on the plot. For example, for 4 dummy rows and an aggressor activation count of 18 (leftmost box in~\figref{fig:utrr}), each dummy row is activated $\lfloor{}(78 - 18*2)/4\rfloor{} = 10$ times.}

\begin{figure}[!ht]
    \centering
    \includegraphics[width=\linewidth]{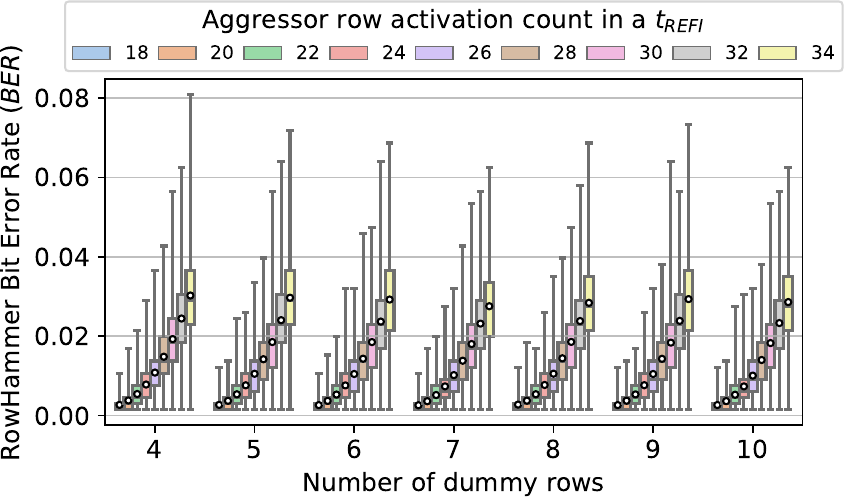}
    \caption{{RowHammer \gls{ber} distribution across all tested rows for the specialized \revcommon{access} pattern with different number of dummy rows (x-axis) and different aggressor row activation counts (different colored boxes).}}
    
    \label{fig:utrr}
\end{figure}

\atbcrcomment{3}{We do not see bitflips for X=0, 1, 2, 3.}{We make \param{\omcr{3}{four}} key observations from Fig.~\ref{fig:utrr}. \atbcr{3}{First, our specialized access patterns can induce RowHammer errors with a reasonably high \gls{ber}.} \omcr{3}{Second}, the \revcommon{access} pattern needs to activate at least 4 dummy rows to bypass the TRR mechanism \atbcr{3}{(i.e., \gls{ber} is 0 for x = 1, 2, and 3)}. \omcr{3}{Third}, the number of dummy rows does \emph{not} significantly affect the distribution of the bit error rate. For example, the mean bit error rate varies by {\param{0.003} \revb{pp}} between the largest (\param{4} dummy rows) and the smallest (\param{7} dummy rows) value at an aggressor activation count of \param{34}. \omcr{3}{Fourth}, the number of bitflips per row increases as the aggressor activation count increases. For example, the mean bit error rate increases by \param{2.79}$\times{}$, \param{6.72}$\times{}$, and \param{10.28}$\times{}$ as the aggressor activation count increases from 18 to 24, 30, and 34, respectively, when the number of dummy rows is 8.}

\take{A specialized \revcommon{access} pattern that bypasses the undocumented TRR mechanism in HBM2 chips can induce many RowHammer bitflips.}
\label{take:trr-sucks}



\section{Implications on Future Read Disturbance Attacks and Defenses}
\label{sec:implications}
We describe and analyze the important implications of our observations 
on future read disturbance attacks and defenses.

\subsection{Read Disturbance Attacks}
Observation~\ref{obs:ber-high} and Takeaways~\ref{take:rowhammer-change-channel},~\ref{take:taggon}, and~\ref{take:there-is-trr} have the following \param{\omcr{3}{four}} implications for future read disturbance attacks on HBM2 chips. First, the maximum \gls{ber} (\param{247} bitflips in a row of 8192 bits) we observe across tested chips exceeds the correction capabilities of widely used error correcting codes (ECC), such as SECDED~\cite{mukherjee2008architecture, nvidia2016pascal,gurumurthi2021hbm3,nvidia-ampere-indepth}\atbcrcomment{3}{SECDED is used in real HBM2 products from NVIDIA and AMD. Chipkill does not have a 1-1 equivalent in HBM as there is a single chip. You can implement single symbol correction within a single cache line (e.g., up to 16-bit correction in one of the 16 consecutive bits in a cache line) but RH errors go beyond this symbol boundary easily.} which can detect two bitflips and correct one bitflip in a codeword (Observation~\ref{obs:ber-high}).\footnote{\atbcr{3}{Single symbol correcting codes (e.g., Chipkill)~\cite{amd2013sddc,yeleswarapu2020addressing, chen1996symbol} are widely used in DDRx server systems. A prior work~\cite{gurumurthi2021hbm3} proposes implementing \atbcr{4}{a Chipkill-like} scheme for HBM3 where 1) the HBM3 chip is carefully architected to ensure a hard fault in a DRAM component manifests as errors \emph{only} within an \emph{isolation boundary} (e.g., 16 consecutive bits), 2) a single symbol correcting code is used where each symbol corresponds to an isolation boundary. Such a scheme \emph{alone} is likely \emph{not} a good read disturbance countermeasure as read disturbance errors typically appear in multiple isolation boundaries.}} \param{247} bitflips in a row are already sufficient to induce uncorrectable bitflips in multiple 64-bit words in the same DRAM row (by pigeonhole principle) and are likely to induce undetectable bitflips in at least one 64-bit word.

\noindent
\revlabel{Rev.C-C1}\textbf{\revc{Analysis of the effectiveness of ECC.}}
{We investigate the distribution of word-level RowHammer bitflips (i.e., bitflips that occur in all non-overlapping consecutive 64 bits) in Chip 4. \figref{fig:ecc-analysis} plots the number of words \revc{that exhibit} at least one bitflip (out of all 18M tested words) on the y-axis over \atbcr{3}{1, 2, ..., and more than 7} bitflips in a word depicted on the x-axis for different data patterns (different bars). The word counts across clusters of bars are non-overlapping; for example, the middle cluster depicts the number of words with exactly two bitflips.}

\begin{figure}[!h]
    \centering
    \includegraphics[width=\linewidth]{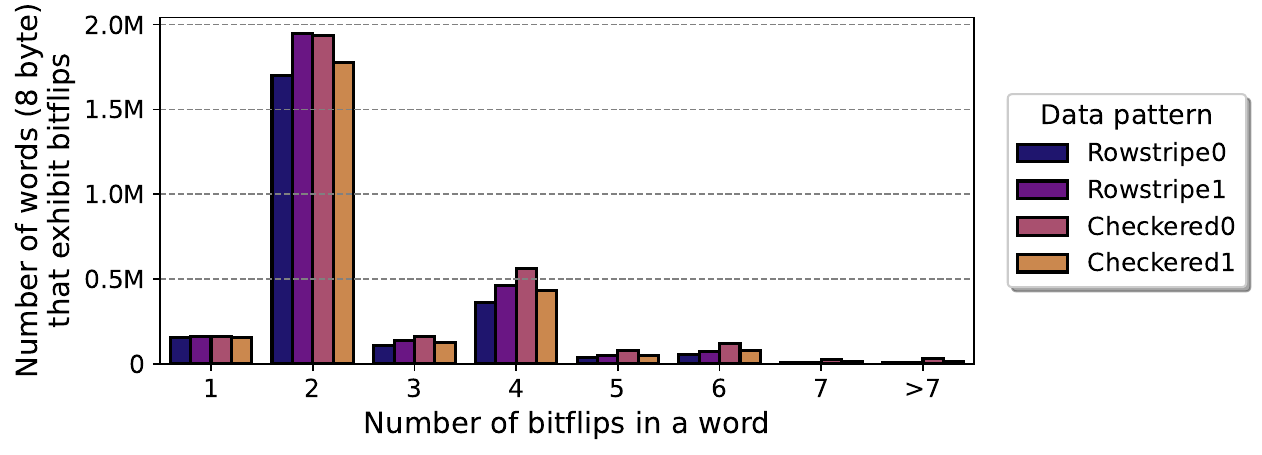}
    \caption{{Number of words (y-axis) that \atbcr{3}{exhibit 1, 2, ..., and more than 7} bitflips (x-axis) in Chip 4.}}
    \label{fig:ecc-analysis}
\end{figure}

{We make two observations. First, the number of unique words with more than two bitflips is large. We observe 974'935 words with more than two bitflips when we use the Checkered0 data pattern. These bitflips would \emph{not} be detected by SECDED ECC. Second, most words with at least one bitflip actually have more than one bitflip. In other words, if RowHammer bitflips can be induced in a word, it is very likely that more than one bit location in the word will experience errors. Simple SECDED ECC \emph{cannot} correct such bitflips.}

{We find that an HBM word can have 16 bitflips (not shown in the figure) in Chip 4. \omcr{3}{A} (7,4) Hamming code~\cite{hamming1950error} could correct such bitflips at very large DRAM storage overheads (75\%, 3 parity bits for every 4 data bits). Thus, relying on ECC alone to prevent RowHammer bitflips in HBM2 is a very expensive solution.}
The bitflip distributions indicate that attackers could exploit RowHammer bitflips to escalate privilege and leak security-critical and secret data in HBM2 chips, \omcr{3}{similarly to} real RowHammer attacks on DDR4-based computing systems. Even if an HBM2 chip is highly RowHammer resilient (i.e., has small mean \gls{ber} across its rows), malicious parties could exploit RowHammer bitflips to practically increase the rate of correctable or detectable bitflips. This could reduce the lifetime of modern HBM-based systems (e.g., GPUs) by accelerating the rate of memory page retirements~\cite{nvidia2019dynamic} beyond design-time estimates and exacerbate system maintenance costs.

Second, a RowHammer attack could target the most-RowHammer-vulnerable HBM2 channel to reduce the time it spends on i) \emph{preparing} for an attack, by finding exploitable RowHammer bitflips faster (i.e., by accelerating memory templating), and ii) \emph{performing} the attack, by benefitting from a small \gls{hcfirst} value (Takeaway~\ref{take:rowhammer-change-channel}). Third, an attacker could keep the aggressor row on for longer by executing specialized access patterns (as demonstrated by RowPress in a real DDR4-based system~\cite{luo2023rowpress}) to benefit from the increased \gls{ber} and reduced \gls{hcfirst} when \omcr{3}{an} aggressor row \omcr{3}{is} kept open for longer (Takeaway~\ref{take:taggon}). Fourth, the RowHammer attack must uncover and take into account the functionality of the undocumented TRR mechanism in addition to the functionality of the documented TRR mode~\cite{jedec2021hbm} to come up with an effective access pattern that bypasses all RowHammer defense mechanisms in an HBM2-based system (Takeaway~\ref{take:there-is-trr}). However, the attackers can also benefit from the undocumented TRR mechanism. \atbcr{3}{Victim row refresh operations} could be used as a near aggressor row activation, carrying over the read disturbance effects of the far aggressor row to the victim row in a HalfDouble access pattern~\cite{kogler2022half}.

\subsection{Read Disturbance Defenses}

Takeaways~\ref{take:rowhammer-change-channel},~\ref{take:subarray}, and~\ref{take:trr-sucks} have the following \param{\omcr{3}{two}} implications for future read disturbance \atbcr{3}{defenses} on HBM2 chips.
First, a defense mechanism can adapt to the heterogeneous distribution of the RowHammer and RowPress vulnerabilit\omcr{3}{ies} across channels and subarrays, which may allow the defense mechanism to more efficiently prevent read disturbance bitflips (Takeaways~\ref{take:rowhammer-change-channel} and~\ref{take:subarray}). 
Second, HBM2 memory controller designers likely need to implement other read disturbance defense mechanisms (e.g.,~\mitigatingRowHammerAllCitations{}) in their designs because designers cannot rely on the undocumented TRR mechanism to mitigate read disturbance bitflips, as \revcommon{it is easily bypassed with a \revcommon{specialized} RowHammer access pattern} (Takeaway~\ref{take:trr-sucks}).

\section{Related Work}

{We present the first detailed experimental characterization of the read disturbance vulnerability (RowHammer and RowPress) in modern HBM2 DRAM chips.} 

\noindent
\textbf{HBM2 RowHammer Characterization~\cite{olgun2023hbm,nam2023xray}.} A prior work~\cite{olgun2023hbm} experimentally characterizes the RowHammer vulnerability in an HBM2 DRAM chip. \dsnadd{Another work~\cite{nam2023xray} studies the internal DRAM structure by analyzing RowHammer error characteristics of two HBM2 chips.} Our work presents a detailed experimental characterization of both RowHammer and RowPress using six HBM2 chips. In addition to analyzing the spatial variation of \atbcr{3}{read disturbance vulnerability}, we analyze the hammer counts needed to induce the first 10 bitflips in a row. We uncover entirely the inner workings of the read disturbance defense mechanism in an HBM2 chip and demonstrate RowHammer access patterns that bypass this defense mechanism. {Our new analyses and results lead to completely new observations (Observations 2, 4, 7, 8, 11, \revcommon{13-23}) and takeaways (Takeaway 1, 2, 5-8) that~\cite{olgun2023hbm} \dsnadd{and~\cite{nam2023xray}} do \emph{not} contain.}

\noindent
\textbf{\omcr{3}{(LP)}DDR3/4 Read Disturbance Characterization~\cite{orosa2021deeper,kim2020revisiting,kim2014flipping, yaglikci2022understanding, lim2017active, park2016statistical, park2016experiments, ryu2017overcoming, yun2018study, lim2018study, luo2023rowpress, lang2023blaster, yaglikci2024svard}.}
These works experimentally demonstrate and analyze new aspects of the read disturbance vulnerability by testing real \omcr{3}{(LP)}DDR3/4 DRAM chips. \atbcr{1}{They} do \emph{not} experimentally analyze real HBM2 chips.

{Besides demonstrating the interaction between the read disturbance vulnerability of an HBM2 chip with unique HBM characteristics 
for the first time, we make new observations (Observations 10, 11, \revcommon{13}, 14, 15, 16) that could also shed light \atbcr{4}{on} the read disturbance vulnerability behavior in (LP)DDRx chips. For example, our observation of the RowHammer bit error rate peaking towards the middle of a subarray (Observation 10) could be \revcommon{widespread} across many different types of DRAM chips. \secref{sec:implications} highlights the importance of our new observations in the form of several key implications for future RowHammer attacks and defenses.}

\noindent
\textbf{Other HBM2 Characterization~\cite{larimi2021understanding, kwon2023temperature, sullivan2021characterizing}.}
These works characterize real HBM2 chips to understand their 1)~soft error resilience characteristics by high-energy neutron beam testing~\cite{sullivan2021characterizing}, 2)~performance and reliability characteristics under reduced supply voltage~\cite{larimi2021understanding}, and 3)~data retention characteristics~\cite{kwon2023temperature}. 

\section{Conclusion}

\atbcr{1}{W}e present the results of our detailed characterization study on the read disturbance (RowHammer and RowPress) vulnerability in six modern HBM2 chips.
\atbcr{3}{Our study leads to 23 observations and 8 takeaways,} which we believe have important implications for future read disturbance attacks and defenses.
We uncover the inner workings of \atbcr{1}{the proprietary read disturbance mitigation mechanism implemented in an HBM2 chip} and develop a practical access pattern that bypasses it and induces read disturbance bitflips. \atbcr{1}{We hope and expect that our findings will lead to a deeper understanding of and new solutions to the read disturbance vulnerabilit\omcr{3}{ies} in HBM-based systems.}


\section*{\atbcr{1}{Acknowledgements}}

\atbcr{1}{We thank the anonymous reviewers of {HPCA 2024 and DSN 2024} for feedback and the SAFARI Research Group members for {constructive} feedback and the stimulating intellectual {environment.} We acknowledge the generous gift funding provided by our industrial partners ({especially} Google, Huawei, Intel, Microsoft), which has been instrumental in enabling the research we have been conducting on read disturbance in DRAM {in particular and memory systems in general.} This work was in part supported by the Google Security and Privacy Research Award and the Microsoft Swiss Joint Research Center.}

\bibliographystyle{IEEEtran}
\bibliography{rh_refs}

\end{document}